\documentclass[letterpaper, 10 pt, conference]{ieeeconf}  

\IEEEoverridecommandlockouts                              

\usepackage{graphics} 
\usepackage{epsfig} 
\usepackage{amsmath} 
\usepackage{amssymb}  

\usepackage{hyperref}
\usepackage{xcolor}
\usepackage{graphicx}
\usepackage{subcaption}

\usepackage{amsthm}
\newtheorem{theorem}{Theorem}

\title{\LARGE \bf
Continuification-based
control of large multiagent 
systems in a ring
}

\author{Gian Carlo Maffettone$^{1}$, Alain Boldini$^{2}$, Mario di Bernardo$^{1,4, *}$, Maurizio Porfiri$^{2, 3, *}$
\thanks{This work has been partially supported by the research grant ``BIOMASS'' from the University of Naples Federico II - ``Finanziamento della Ricerca di Ateneo (FRA) - Linea B'' and by the National Science Foundation Grant No. CMMI-1932187. The authors wish to acknowledge Dr. Davide Fiore (University of Naples Federico II)  for all the insightful comments.}%
\thanks{$^{1}$Scuola Superiore Meridionale, Naples, Italy}%
\thanks{$^{2}$ Center for Urban Science and Progress and Department of Mechanical and Aerospace Engineering, Tandon School of Engineering, New York University, USA}%
\thanks{$^{3}$ Department of Biomedical Engineering, Tandon School of Engineering, New York University, USA}
\thanks{$^{4}$ Department of Electric Engineering and Information Technology, University of Naples Federico II, Naples, Italy}%
\thanks{$^{*}$For correspondence: {\tt\small mario.dibernardo@unina.it, maurizio.porfiri@nyu.edu}}%
}
\usepackage[T1]{fontenc}

\begin{document}

\maketitle
\thispagestyle{empty}
\pagestyle{empty}

\begin{abstract}

In this paper, we propose a method to control large-scale multiagent systems swarming in a ring. Specifically, we use a continuification-based approach that transforms the microscopic, agent-level description of the system dynamics into a macroscopic, continuum-level representation, which we employ to synthesize a control action towards a desired distribution of the agents. The continuum-level control action is then discretized at the agent-level in order to practically implement it. To confirm the effectiveness and the robustness of the proposed approach, we complement theoretical derivations with a series of numerical simulations.
\end{abstract}

\section{Introduction and Background}
A pressing open challenge in control theory is to find methods to steer the collective behavior of large-scale multiagent systems consisting of many dynamical units (or agents) interacting with a given, and possibly time-varying, network topology. Examples of this problem include multirobot systems \cite{Rubenstein2014, Gardi2022, giusti2022distributed}, 
cell populations, \cite{guarino2020balancing, agrawal2019vitro},  
and human networks \cite{calabrese2021spontaneous, Shahal2020a}.
Typically, in these applications, the goal is to control some macroscopic observables of the emerging collective behavior. However, control needs to be practically exerted at the microscopic, individual agent-level. Developing methods that translate macroscopic-level control goals into microscopic-level control actions is fundamental to steer complex multiagent systems towards desired behaviors and close the feedback loop across different scales \cite{diBernardo2020}. 

To describe large-scale systems, mean field approaches are often used in statistical mechanics and physics \cite{kardar2007statistical, kardar2007statistical_fields}. 
Through a mean field approximation, one can obtain a macroscopic description of the emergent behavior of the system in terms of appropriate distributed parameter models, derived from the microscopic ordinary differential equations (ODEs) models describing the agents' dynamics.
Such mean field approaches have been also proposed to control the collective behavior of multiagent systems \cite{Albi2020a, Elamvazhuthi2021, Borzi2020, Kolpas2007}. 
In the applied mathematics community, problems related to the control of crowds, herding, and  flocking agents were also solved by finding a mean field description of the agents of interest, then used to compute an open-loop optimal control strategy \cite{Albi2020a, ascione2022mean}.  
Another methodology recently proposed in the literature is
based on the use of \emph{graphons}, see for example \cite{Gao2020}. 
Other methods based on data and manifold learning are proposed in \cite{lee2020coarse, patsatzis2022data} 
where controllers are developed by projecting the agents' dynamics on a particular manifold of interest. Such an approach may result in non-physical mathematical models, that are not easy to analyse. 

\begin{figure}[tbp]
    \centering
    \includegraphics[width=0.5\textwidth]{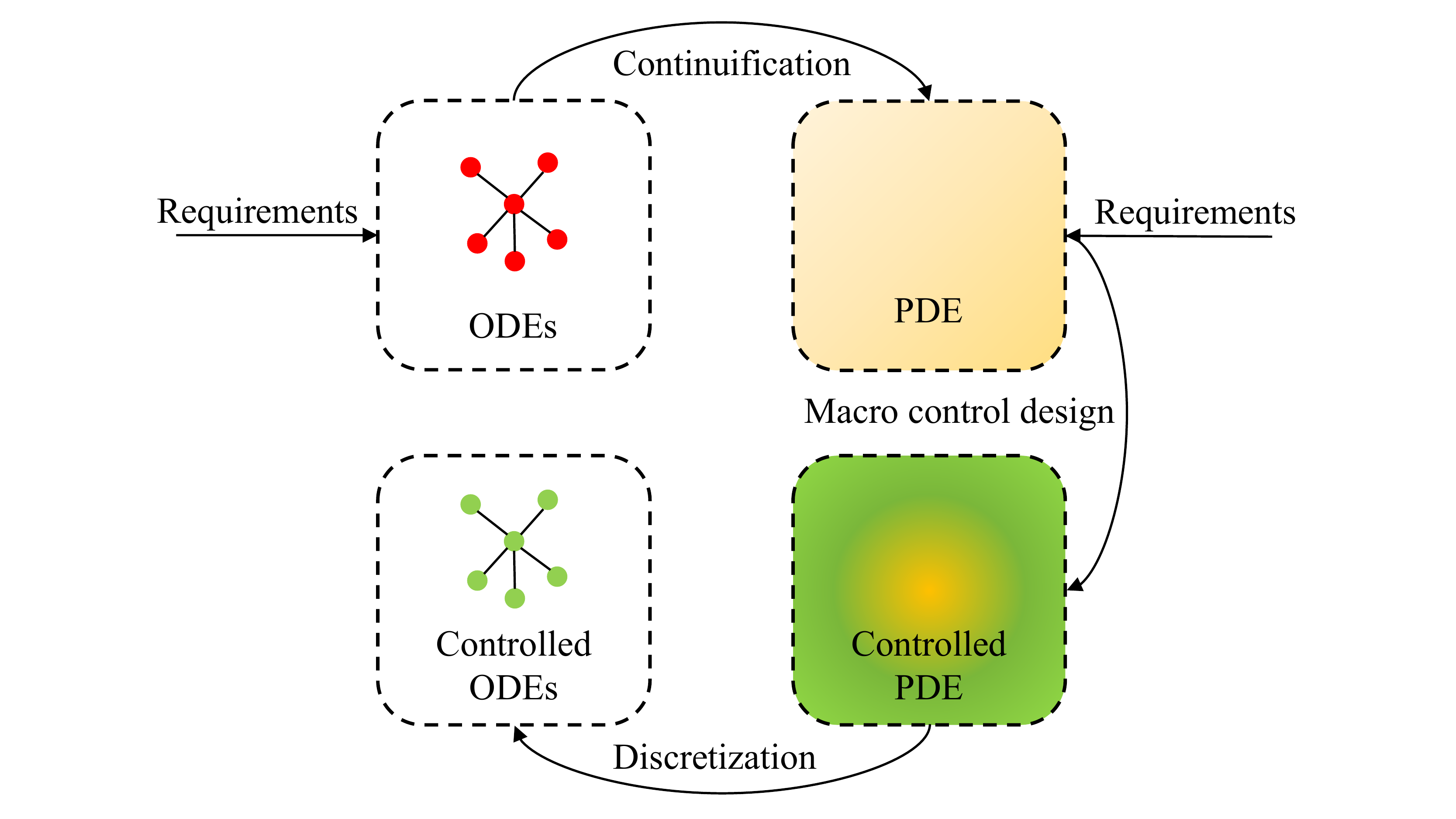}
    \caption{Continuification-based control approach used in this paper (inspired by \cite{nikitin2021continuation}).}
    \label{fig:continuification}
\end{figure}

Inspired from the paradigm proposed in \cite{nikitin2021continuation}
(see fig.~\ref{fig:continuification}), here we adopt a {\em continuification} approach in which a macroscopic model, derived from the agents' dynamics, is used to design a control strategy at the macroscopic level. Such a macroscopic control action is then discretized in order to be deployed on the agents at the microscopic level.
As a representative case of study, we tackle the problem of steering the dynamics of a group of interacting agents on a one-dimensional periodic domain (a ring). Our goal is to control the agents so that they achieve some desired configuration, independently of their interactions (repulsive, attractive, etc.) and their initial configuration. Such a problem has important ramifications in traffic dynamics \cite{Karafyllis2019, Liard2020}, swarming robots
\cite{Elamvazhuthi2021}
and natural systems, including animals' collective motion \cite{Zienkiewicz2018, Abaid2010, DeLellis2020, Aureli2010,Yates2009, leverentz2009asymptotic, bernoff2011primer}, and cell populations \cite{agrawal2019vitro}.

After presenting the microscopic description of the agents' behavior, we derive a macroscopic, partial differential equations (PDE) model for their emergent behavior and we solve the problem of designing a control strategy to achieve a desired agents' configuration. 
We propose a mathematical proof of local convergence at the macroscopic level, and then discretize the control action to obtain the required control inputs acting on the individual agents. Differently from what proposed in \cite{nikitin2021continuation}, the microscopic control inputs are obtained by spatially sampling the macroscopic control action at the agents' positions.
Theoretical derivations are complemented by a set of numerical simulations validating the effectiveness and robustness of the proposed strategy in a number of representative scenarios.

\section{Model and Problem Statement}
\subsection{The model}
Let $\mathcal{X}$ be a group of $N$ identical mobile agents moving on the unit circle $\mathcal{S}=[-\pi,\pi]$.
By making the \emph{kinematic} assumption widely used in the literature \cite{viscek1995} (that is, neglecting acceleration and considering a drag force proportional to the velocity), the dynamics of the $i$-th agent can be expressed as
\begin{equation}
    \dot{x}_i = \sum_{j = 1}^N f\left(\left[x_i,x_j\right]_{\pi}\right) + u_i
    \label{eq:themodel},
\end{equation}
where $x_i$ is the angular position of agent $i$ on $\mathcal{S}$, $[x_i,x_j]_{\pi}=(x_i-x_j+\pi)\mod(2\pi)-\pi$ is the angular distance between agents $i$ and $j$ wrapped on $\mathcal{S}$, $u_i$ is the velocity control input affecting its behavior, and $f:\mathbb{R}\to\mathbb{R}$ is a velocity interaction kernel modeling pairwise interactions between the agents. 


Assuming the number of agents to be sufficiently large, we describe the macroscopic collective behavior emerging from the microscopic agents' dynamics in terms of the density profile of agents on $\mathcal{S}$ at time $t$, $\rho : \mathcal{S} \times \mathbb{R}_{\geq 0} \to \mathbb{R}_{\geq 0}$. This function is such that, when integrated over a subset of the domain $\mathcal{S}$, it returns the number of agents occupying that subset. By definition, we require that $\int_\mathcal{S} \rho(x,t) \,\mathrm{d}x=N$ for all $t$. 

\begin{figure}[tbp]
     \centering
     \begin{subfigure}[b]{0.23\textwidth}
         \centering
         \includegraphics[width=\textwidth]{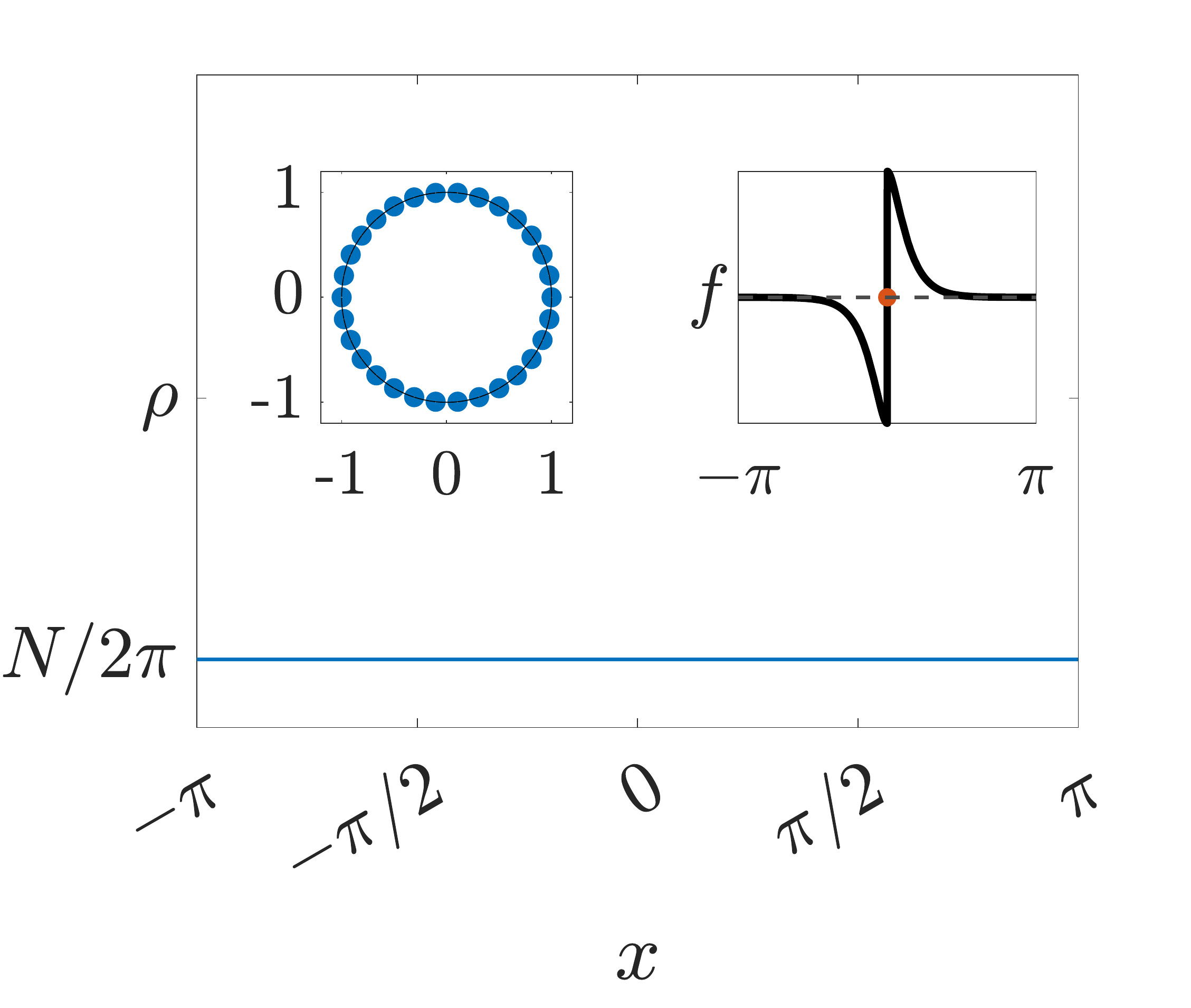}
         \caption{Spreading}
         \label{subfig::f2}
     \end{subfigure}
     \begin{subfigure}[b]{0.23\textwidth}
         \centering
         \includegraphics[width=\textwidth]{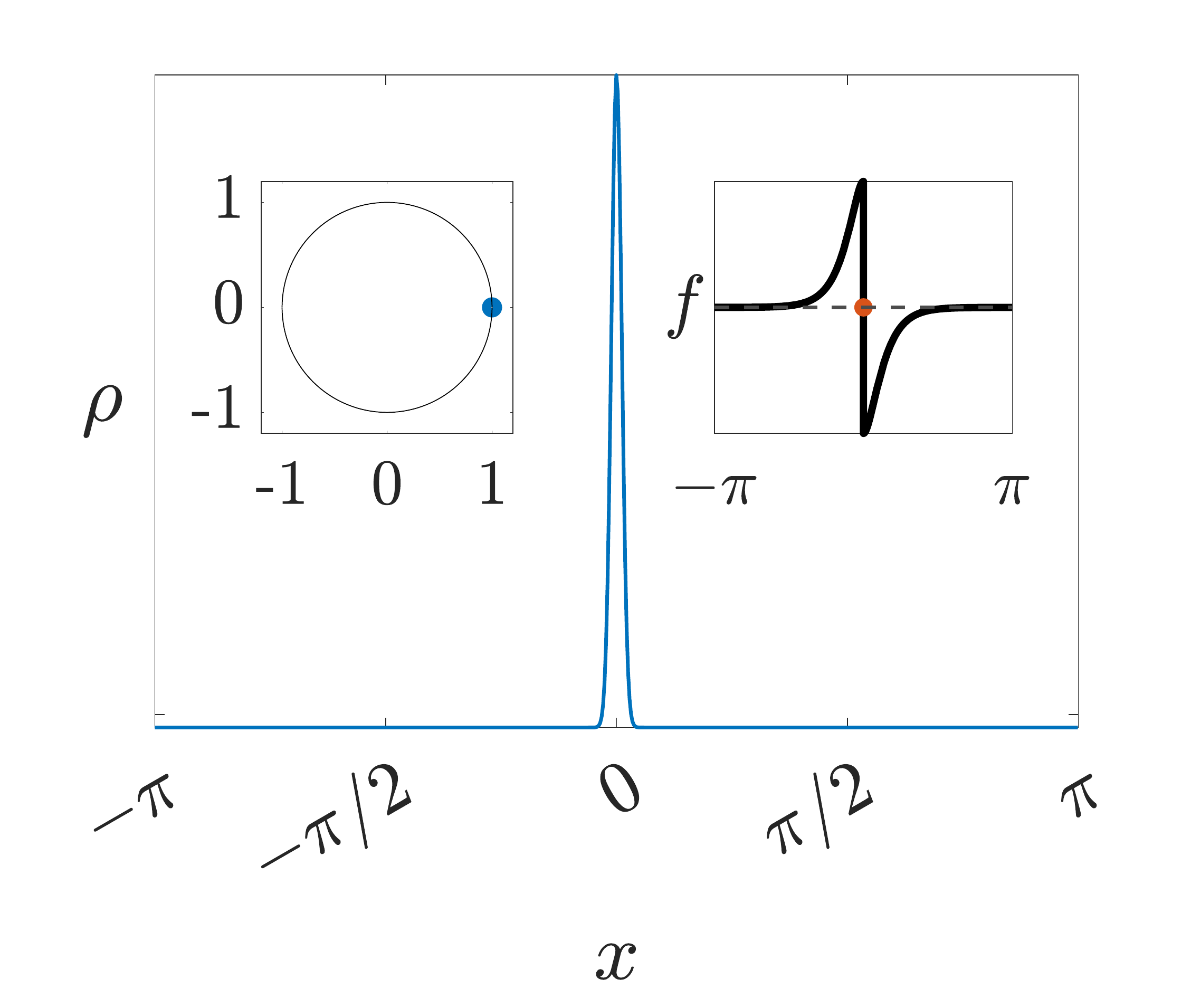}
         \caption{Collapsing}
         \label{subfig::f3}
     \end{subfigure}
     
     \begin{subfigure}[b]{0.23\textwidth}
         \centering
         \includegraphics[width=\textwidth]{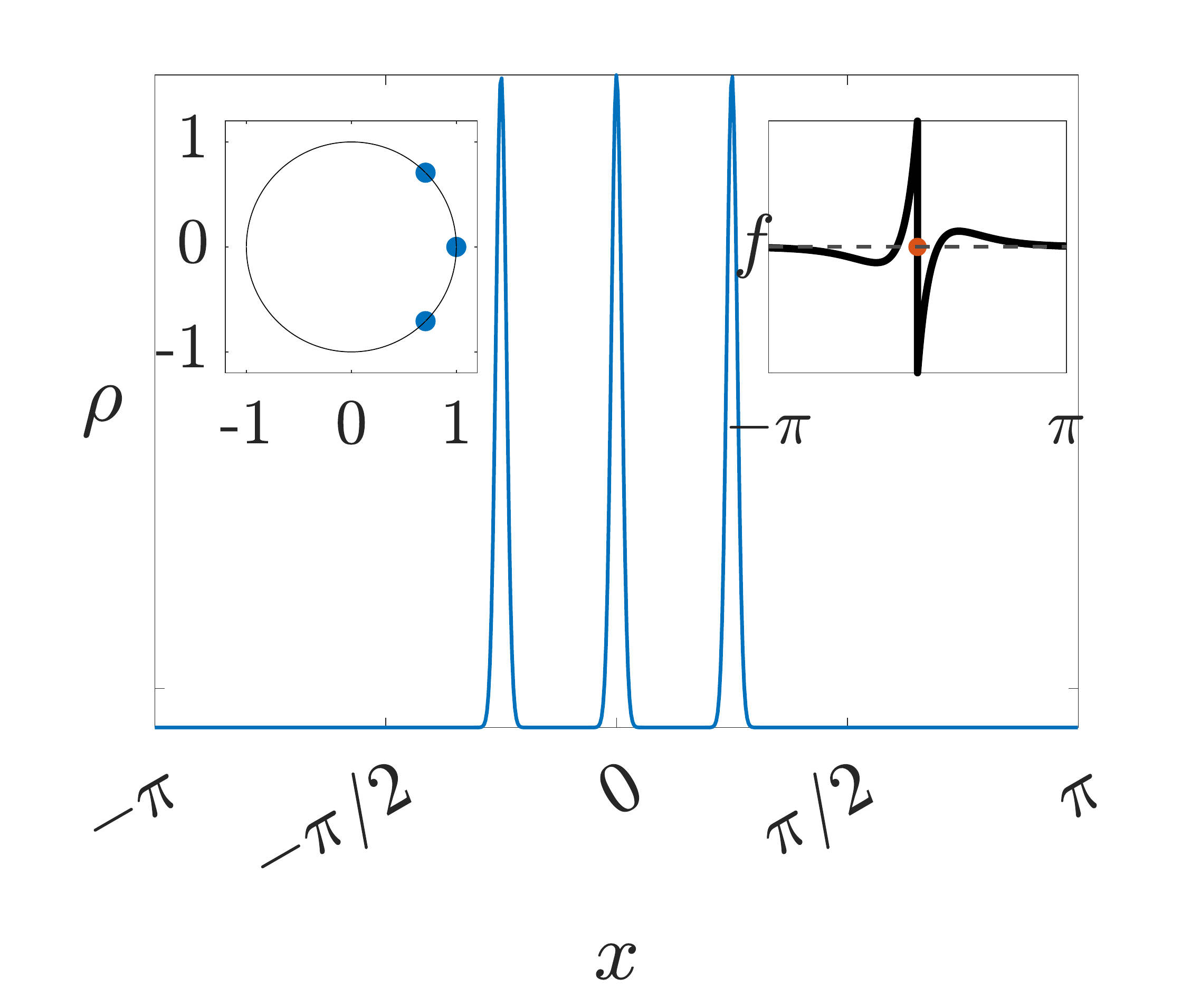}
         \caption{Clustering}
         \label{subfig::f4}
     \end{subfigure}
     \begin{subfigure}[b]{0.23\textwidth}
         \centering
         \includegraphics[width=\textwidth]{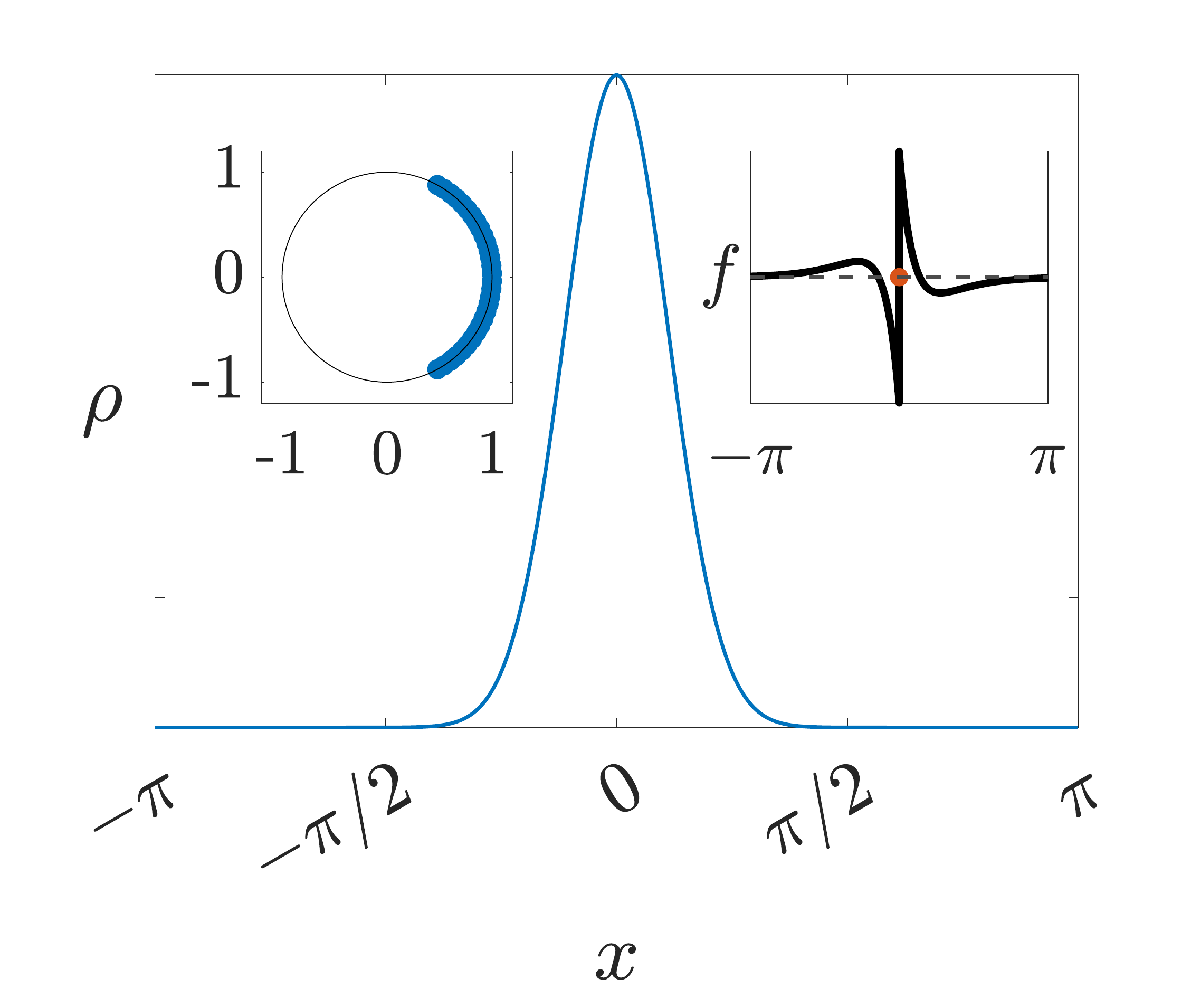}
         \caption{Stable aggregation}
         \label{subfig::f1}
     \end{subfigure}
        \caption{Typical emerging behaviors arising from \eqref{eq:themodel} rotated around the origin. The asymptotic agents' density is shown in each panel together with their distribution on the ring (left inset) and the corresponding interaction kernel (right inset).}
        \label{fig::forces}
\end{figure}

The function $f$ takes the relative angular distance between two agents and returns a velocity. Similar to \cite{leverentz2009asymptotic, bernoff2011primer}, we assume that $f$ is a vanishing odd function,  discontinuous at the origin, where it takes zero value.
As shown in fig.~\ref{fig::forces}, the kernel can take different functional forms that model various types of interactions acting at different ranges. 

In the open-loop scenario, where the control input $u_i$ is set to zero for all the agents, four types of emerging behaviors can occur depending on the initial configuration of the agents and on the different functional form of the interaction kernel (modeling different ranges of attraction and/or repulsion) \cite{leverentz2009asymptotic, bernoff2011primer}: \emph{spreading} (see fig.~\ref{subfig::f2}), \emph{collapsing} (see fig.~\ref{subfig::f3}), \emph{clustering},  (see fig.~\ref{subfig::f4}), or \emph{stable aggregation} (see fig.~\ref{subfig::f1}).    
%
\subsection{Problem statement}
\label{sec:prob_stat}
The problem is to select a set of distributed control inputs $u_i$ acting at the microscopic, agent-level in order for the agents to organize themselves into a desired macroscopic configuration on $\mathcal{S}$. 
Specifically, given some desired periodic smooth density profile, say $\rho^\text{d}(x, t)$, associated with the target agents' configuration, the problem can be reformulated as that of finding a set of distributed control inputs $u_i,\ i=1,2,\dots,N$ in \eqref{eq:themodel} such that
\begin{equation}
    \lim_{t\rightarrow \infty} \Vert{\rho^\text{d}(\cdot, t)}-\rho(\cdot, t)\Vert_2 =0,
    \label{eq:controlgoal}
\end{equation} 
for agents starting from an initial configuration $x_i(0)=x_{i0}, \ i=1,2,\ldots,N$ that is proximal to the one prescribed by $\rho^{\mathrm{d}}(x,0)$ -- here, $\Vert\cdot\Vert_2$ is the $L^2$ norm in $\mathcal{S}$.


\section{Control Design}
We assume the multiagent system of interest to be described by a large set of $N$ coupled ordinary differential equations (ODEs), and adopt an approach based on \emph{continuification}\footnote{Here, we use the term \emph{continuification} instead of continuation, used in \cite{nikitin2021continuation}, to distinguish this procedure from the parametric continuation of dynamical systems.} (or continuation) \cite{nikitin2021continuation}. We describe next how each of the four steps depicted in fig.~\ref{fig:continuification} can be implemented to solve the problem of interest. 

\subsection{Continuification}
Following \cite{leverentz2009asymptotic, bernoff2011primer, Bodnar2005a}, we can derive the macroscopic model describing the open-loop dynamics of \eqref{eq:themodel} when $u_i=0$, analogous to a mass conservation law, 
\begin{equation}
    \rho_t(x,t)  + \left[\rho(x,t) V(x,t)\right]_x = 0, \ \forall x \in  [-\pi, \pi], \ \forall t\geq 0
    \label{eq::macro_model},
\end{equation}
where $V$ is the velocity field characterising the advection term of \eqref{eq::macro_model}, which can be expressed as
\begin{equation}
    V(x, t) = \int_{-\pi}^\pi f\left(\left[x,y\right]_\pi\right)\rho(y, t)\,\mathrm{d}y = (f * \rho) (x, t)
    \label{eq:V},
\end{equation}
where \textquotedblleft{}$*$\textquotedblright{} is the circular convolution operator. For the problem to be well-posed, we shall impose boundary and initial conditions, which read
\begin{eqnarray}
    & &\rho(-\pi,t) = \rho(\pi,t), \quad \forall t\geq 0 \label{eq::periodicity}\\
    & &\rho(x, 0) = \rho^0(x), \quad \forall x \in  [-\pi, \pi],
\end{eqnarray}
where $(\cdot)_t$ and $(\cdot)_x$ denote time and spatial partial derivatives, respectively.
Importantly, for a periodic $\rho$ in $\mathcal{S}$, such as the one we are searching for, the velocity will also be periodic by construction. Hence, the flux is also periodic and the total mass  is constant in time. Specifically, integrating \eqref{eq::macro_model} and using \eqref{eq::periodicity}, we establish that $\int_{\mathcal{S}}\rho_t(x, t)\,\mathrm{d}x = 0$.

 

\subsection{Macroscopic control design}
\label{sec:macro_control_design}
To achieve asymptotic convergence, we consider the addition to \eqref{eq::macro_model} of a control input $q$, representing a mass source/sink term\footnote{Clearly, the control action cannot change the total mass of the system: the choice of writing $q$ as a mass source/sink is only a matter of simplicity of the derivations, and we will ultimately incorporate this term as a control input on the velocity.}.

The resulting closed-loop macroscopic model is
\begin{equation}
    \rho_t(x,t)  + \left[\rho(x,t) V(x,t)\right]_x = q(x, t)
    \label{eq:controlled_model}.
\end{equation}

\begin{figure*}
     \centering
     \begin{subfigure}[b]{0.24\textwidth}
         \centering
         \includegraphics[width=\textwidth]{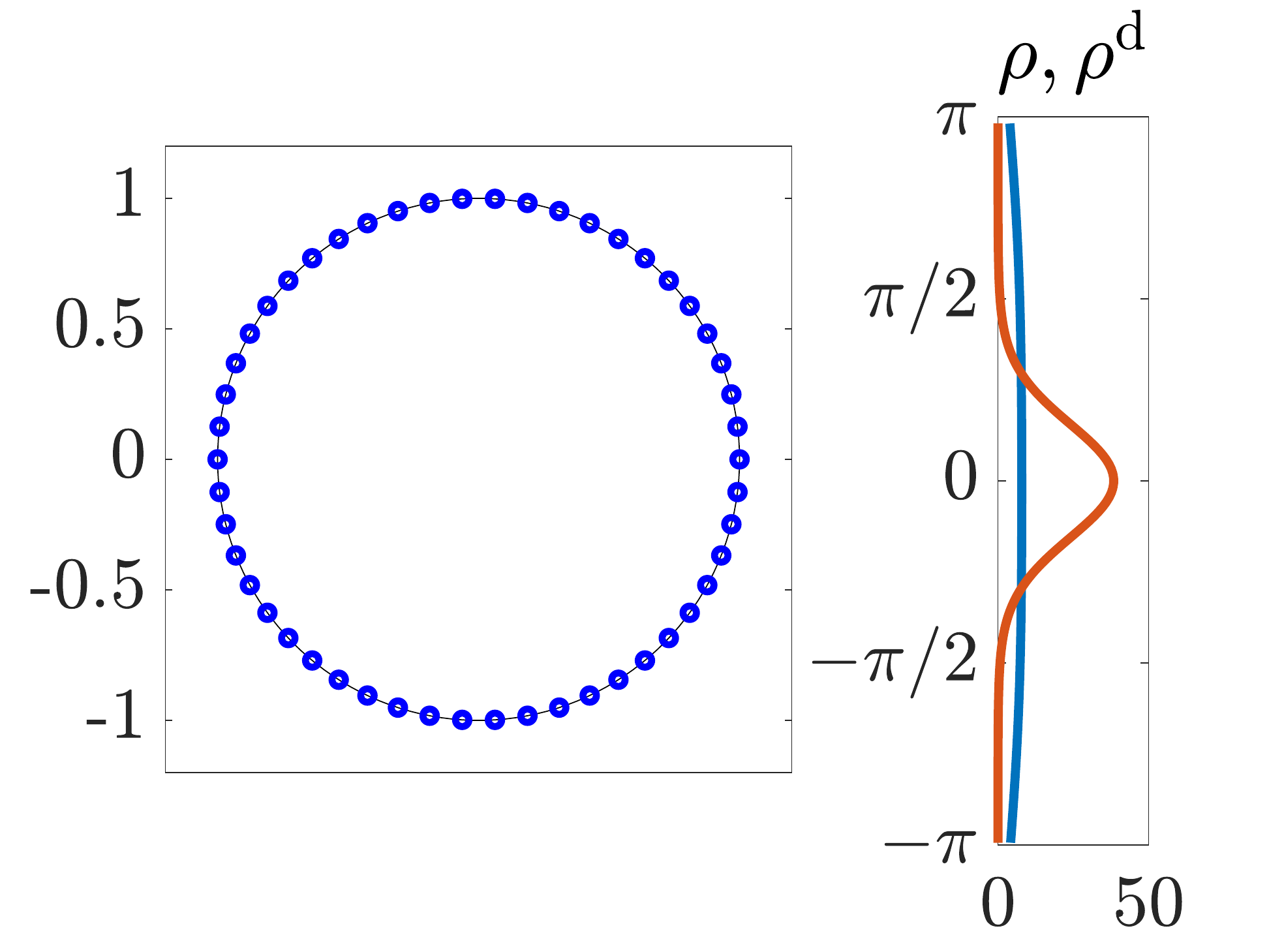}
         \caption{$t = 0$}
         \label{subfig::t0_monompodal_reg}
     \end{subfigure}
     \begin{subfigure}[b]{0.24\textwidth}
         \centering
         \includegraphics[width=\textwidth]{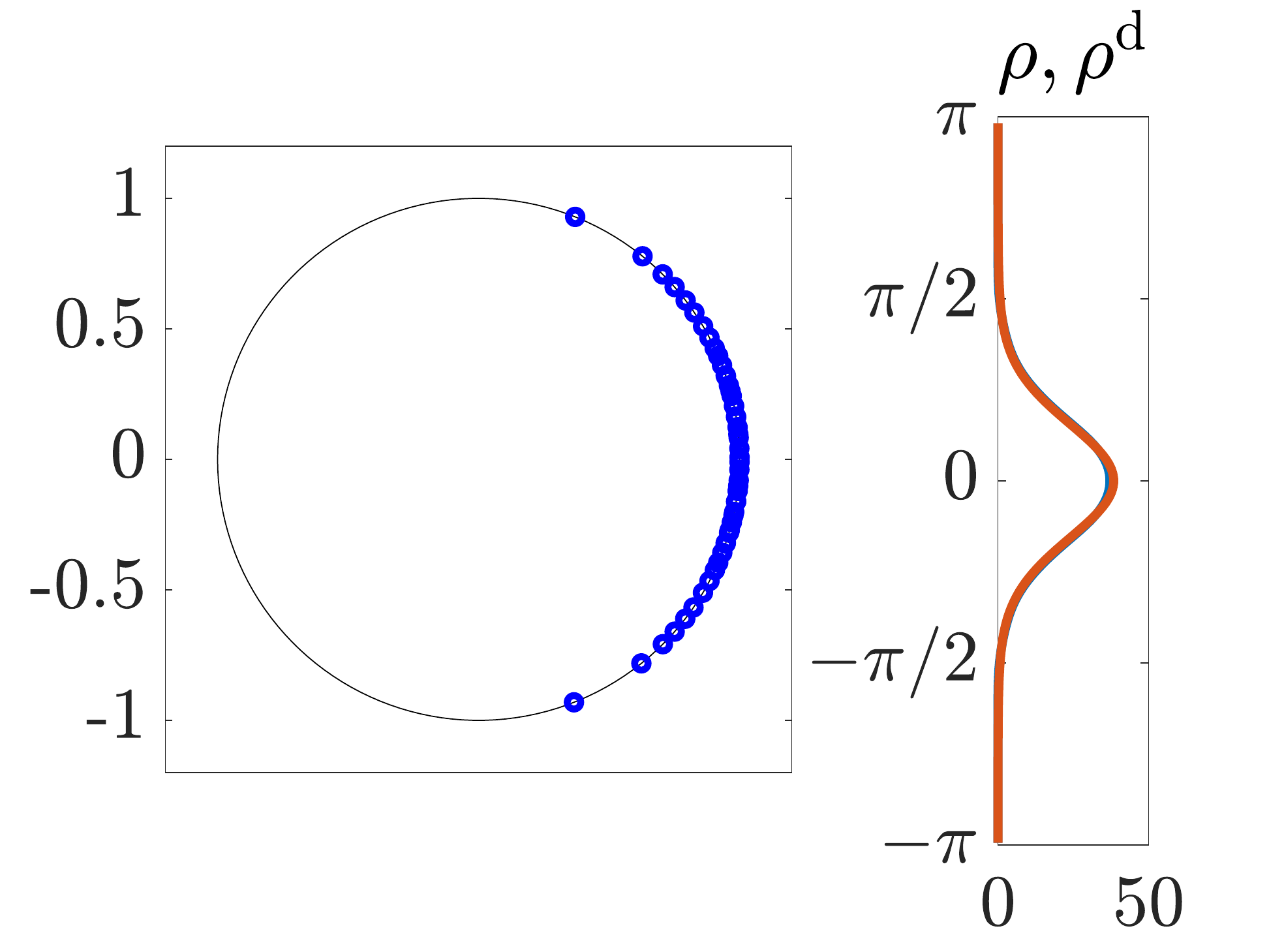}
         \caption{$t = 3$}
         \label{subfig::tf_monompodal_reg}
     \end{subfigure}
     \begin{subfigure}[b]{0.24\textwidth}
         \centering
         \includegraphics[width=\textwidth]{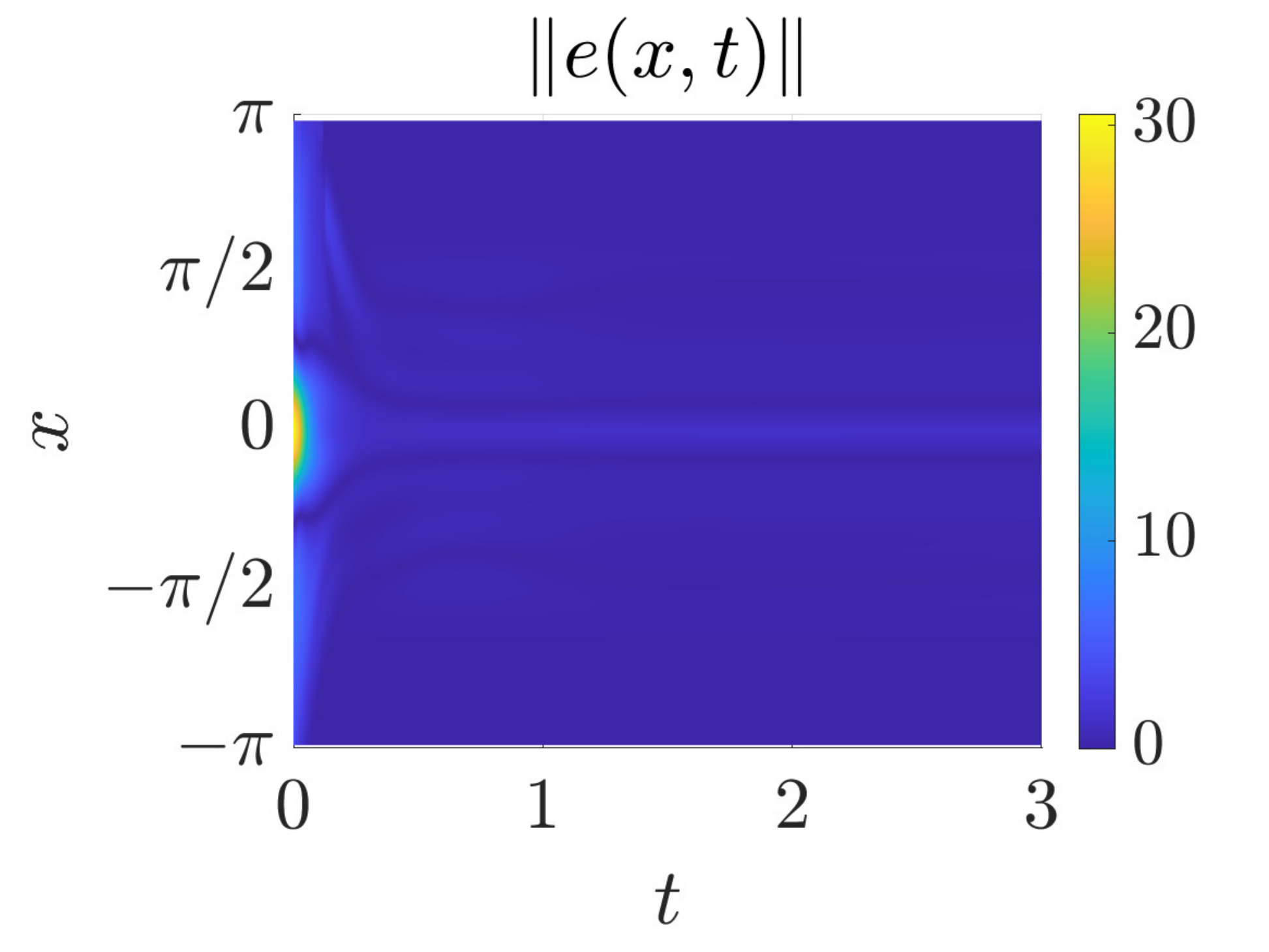}
         \caption{$\Vert e(x, t)\Vert$}
         \label{subfig::e_x_t_monomodal_reg}
     \end{subfigure}
     \begin{subfigure}[b]{0.24\textwidth}
         \centering
         \includegraphics[width=\textwidth]{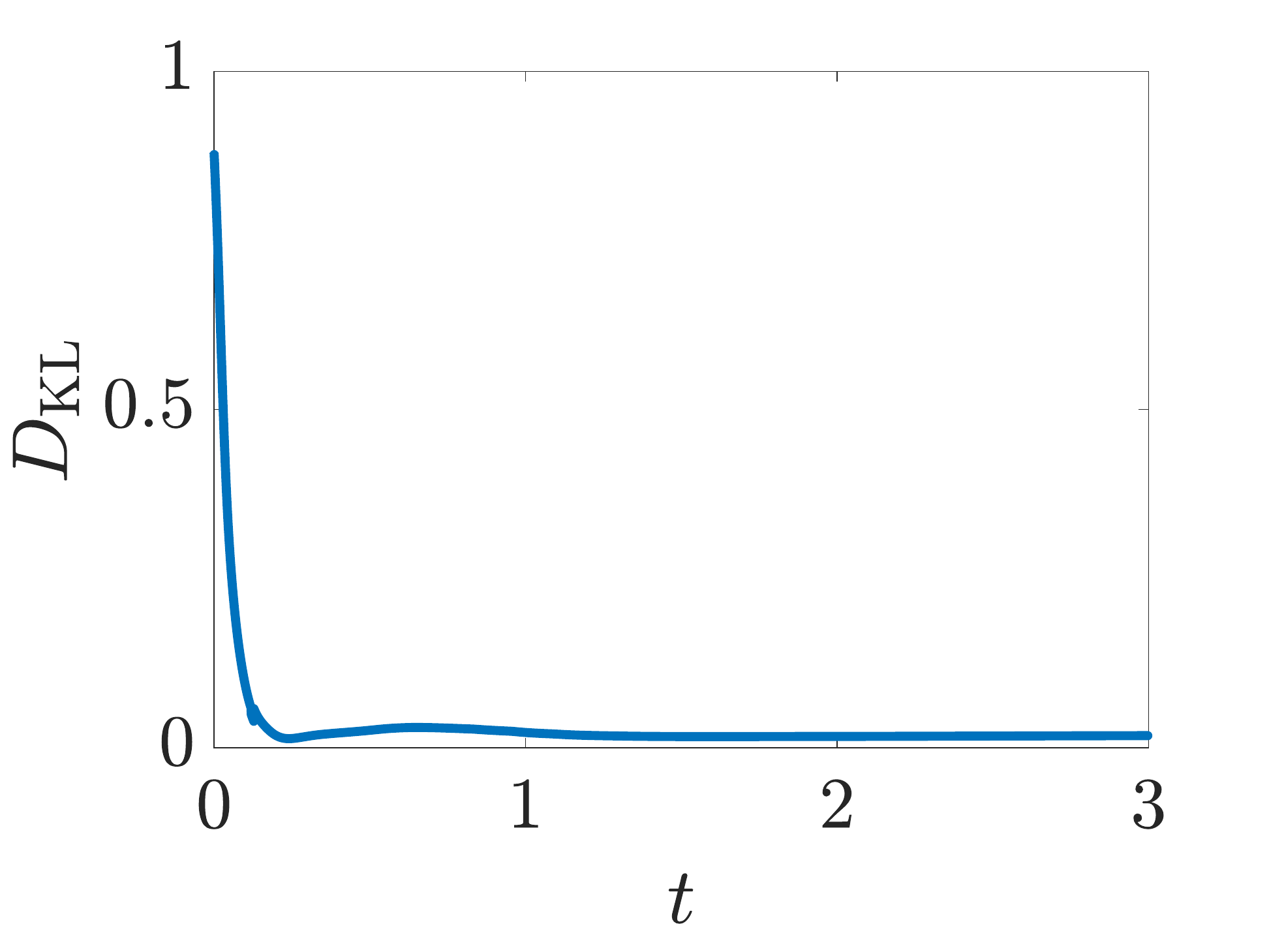}
         \caption{$D_{\mathrm{KL}}(t)$}
         \label{subfig::D_KL_monompodal_reg}
     \end{subfigure}
        \caption{Regulation to a monomodal distribution: (a) initial and (b) final configuration of the agents with their associated density (in blue) and desired density (in orange); (c) time and space evolution of the norm of the error function; (d) KL divergence between the desired normalised density and the normalised density.}
        \label{fig::monompodal_reg}
\end{figure*}
\begin{figure}
    \centering
    \includegraphics[width=0.4\textwidth]{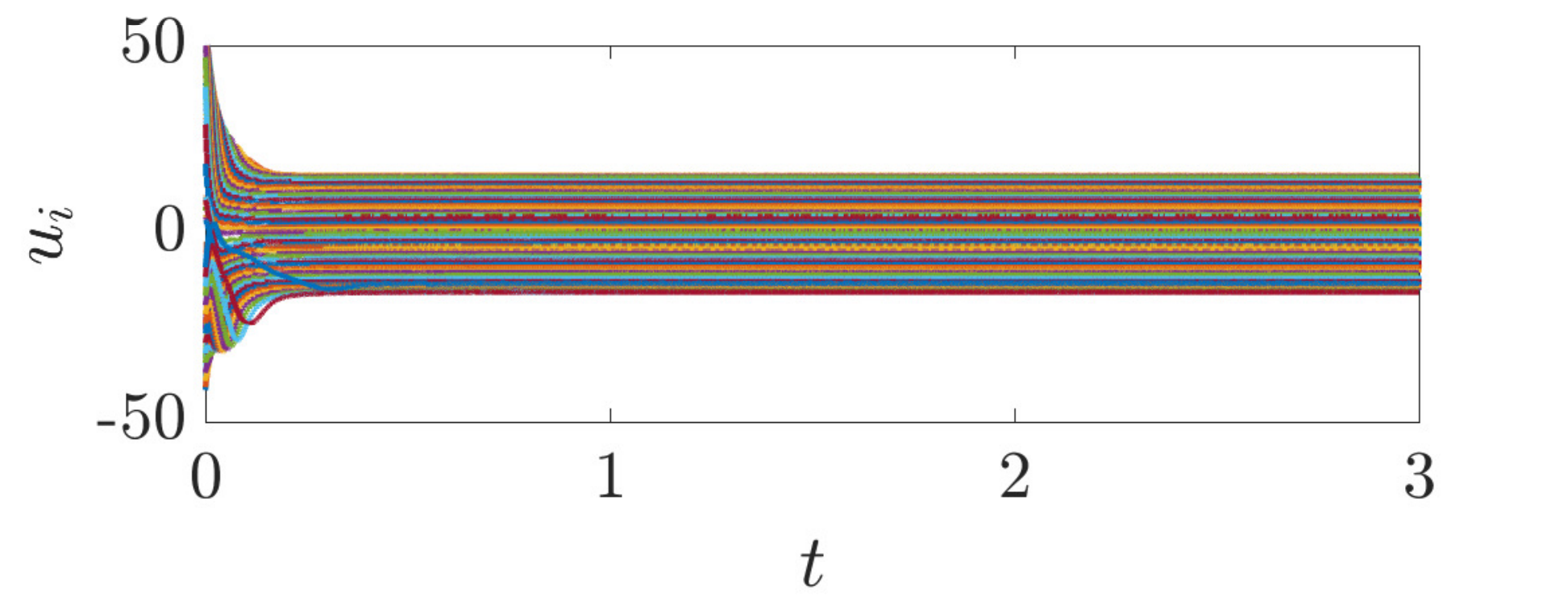}
    \caption{Control inputs computed as in \eqref{eq:u_i}  acting on the agents for the monomodal regulation.}
    \label{fig:control_input_monomodal}
\end{figure}


We design $q$ by assuming the desired density $\rho^\mathrm{d}$ fulfills the following reference dynamics:
\begin{equation}
    \rho^\mathrm{d}_t(x,t) + \left[\rho^\mathrm{d}(x,t)V^\mathrm{d}(x, t)\right]_x = 0
    \label{eq:ref_model},
\end{equation}
where we define
\begin{align}
    V^\text{d}(x, t) = \int_{-\pi}^\pi f\left(\left[x,y\right]_\pi\right)\rho^\text{d}(y, t)\,\mathrm{d}y = (f * \rho^\text{d}) (x, t).
\end{align}
Equation \eqref{eq:ref_model} is a mass conservation law, assumed to fulfil periodic boundary conditions similar to those given in \eqref{eq::periodicity} for \eqref{eq::macro_model}. 

Let $e(x,t)$ be the mismatch between the agents' density and the reference density, defined as
$e(x, t) := \rho^\text{d} (x, t) - \rho (x, t)$.
We set 
\begin{multline}
    q(x, t) = K_\mathrm{p}e(x, t) - \left[e(x, t)V^\mathrm{d}(x, t)\right]_x \\-\left[ \rho^\mathrm{d}(x, t)V^\mathrm{e}(x, t)\right]_x
    \label{eq::q},
\end{multline}
where $K_\text{p}$ is a positive constant gain and we define
\begin{align}
    V^\text{e}(x, t) = \int_{-\pi}^\pi f\left(\left[x,y\right]_\pi\right)e(y, t)\,\mathrm{d}y = (f * e) (x, t).
\end{align}


We can now prove the following theorem.

\begin{theorem} The choice of $q$ as in \eqref{eq::q} guarantees that the closed-loop macroscopic dynamics \eqref{eq:controlled_model} locally asymptotically converges to $\rho^\mathrm{d}(x,t)$ as in \eqref{eq:controlgoal}.
\end{theorem}

\begin{proof}
Using \eqref{eq:ref_model} and and the proposed expression for $q$ in \eqref{eq::q}, yields \eqref{eq:controlled_model} to
\begin{align}
    e_t(x, t) + \left[e(x,t)\left(-V^\mathrm{e}(x, t)\right)\right]_x = -K_\text{p}e(x, t)
     \label{eq::final_err}, 
\end{align}
subject to the boundary and initial conditions
\begin{eqnarray}
     &e(-\pi,t) = e(\pi,t), \quad \forall \ t\geq 0,\\
    &e(x, 0) = \rho^\mathrm{d}(x, 0) - \rho^0(x), \quad \forall \ x \in  [-\pi, \pi].
\end{eqnarray}
We remark that $V = V^\mathrm{d}-V^\mathrm{e}$.
Choosing as a candidate Lyapunov function the square of the $L^2$ norm of the error, $\Vert{e}\Vert^2=\int_{-\pi}^{\pi}e^2(x,t)\mathrm{d}x$, we obtain 
\begin{multline}\label{eq:Lyap}
        \left(\Vert{e}\Vert_2^2\right)_t=2 \int_{-\pi}^{\pi}e(x,t)e_t(x,t)\mathrm{d}x=-2K_p\Vert{e}\Vert_2^2+\Psi
\end{multline}
where 
\begin{multline}
 \Psi=2\int_{-\pi}^{\pi}e(x,t)\left(e(x,t)V^{\mathrm{e}}(x,t)\right)_x\mathrm{d}x= \\ 
 \int_{-\pi}^{\pi}2e^2(x,t)V^{\mathrm{e}}_x(x,t)+\left(e^2(x,t)\right)_xV^{\mathrm{e}}(x,t)\mathrm{d}x
\end{multline}
Integrating by parts the second integrand in $\Psi$ and taking into account the periodicity condition, we can write  
%
\begin{equation}
     \Psi=\int_{-\pi}^{\pi}e^2(x,t)V^{\mathrm{e}}_x(x,t)\mathrm{d}x.
     \label{eq:simpl}
\end{equation}

Now, the spatial derivative of the velocity can be written as  $V^e_x(x,t)=(f_x*e)(x,t)$\footnote{Although $f$ is discontinuous at the origin, its derivative is well-defined and continuous}, whose $\infty$-norm can be bounded using Young's inequality \cite{young1912multiplication} as follows: 
\begin{equation}\label{eq:young}
    \Vert V^{\mathrm{e}}_x\Vert_{\infty}\leq \Vert f_x\Vert_{2}\Vert e\Vert_{2},
\end{equation}
where we note that $\Vert f_x\Vert_{2}$ is independent of time. By using \eqref{eq:simpl} in \eqref{eq:Lyap} along with the bound in \eqref{eq:young}, we establish the following inequality:
    \begin{equation}
    \left(\Vert{e}\Vert_2^2\right)_t\leq (-2K_p+\Vert f_x\Vert_{2}\Vert{e}\Vert_2)\Vert{e}\Vert_2^2.
    \end{equation}
For any $\gamma>0$, if $\Vert{e}\Vert_2<\gamma$, then the error will approach zero for $K_p$ sufficiently large.

\end{proof}

\begin{figure*}
     \centering
     \begin{subfigure}[b]{0.24\textwidth}
         \centering
         \includegraphics[width=\textwidth]{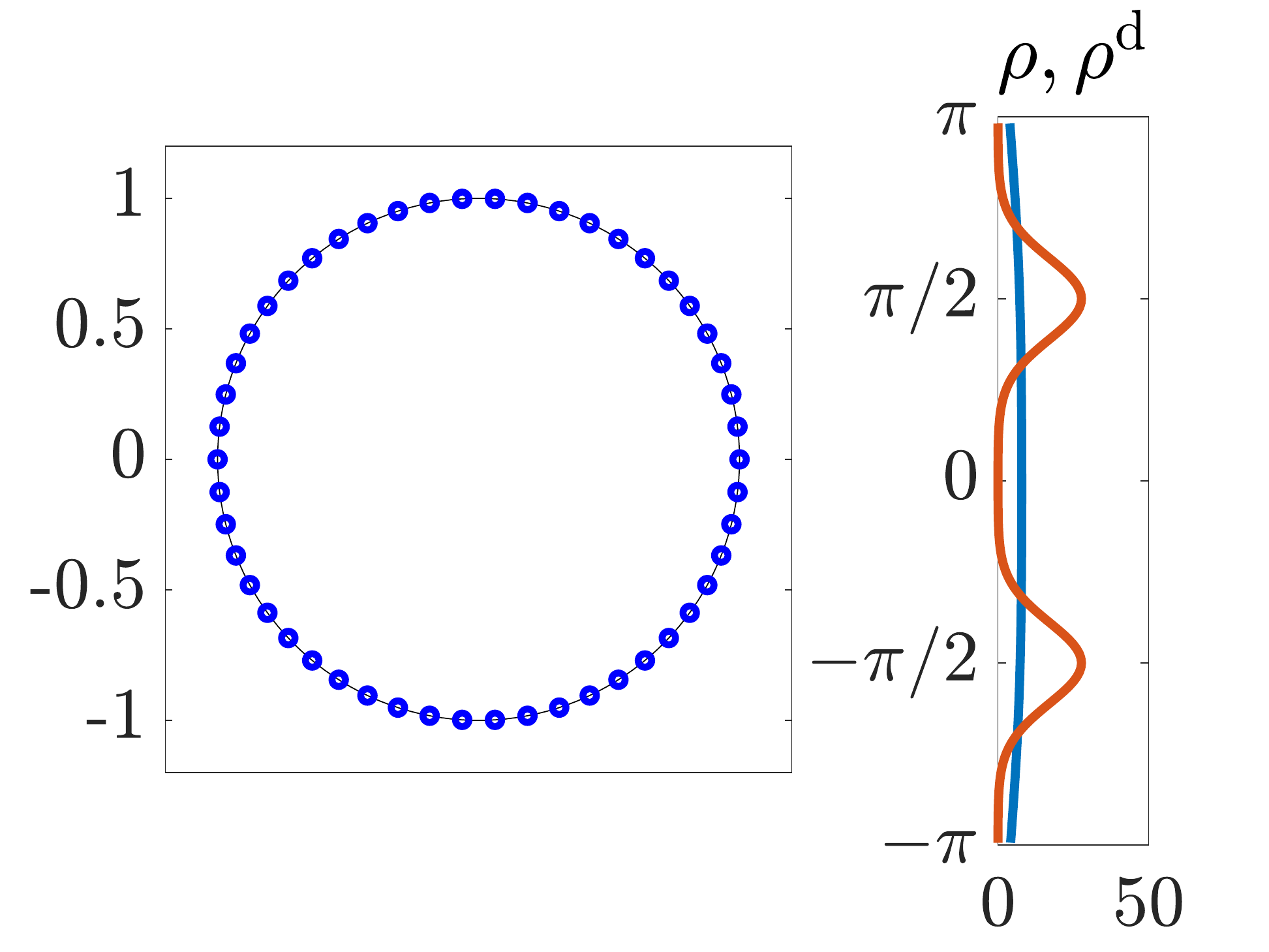}
         \caption{$t = 0$}
         \label{subfig::t0_bimpodal_reg}
     \end{subfigure}
     \begin{subfigure}[b]{0.24\textwidth}
         \centering
         \includegraphics[width=\textwidth]{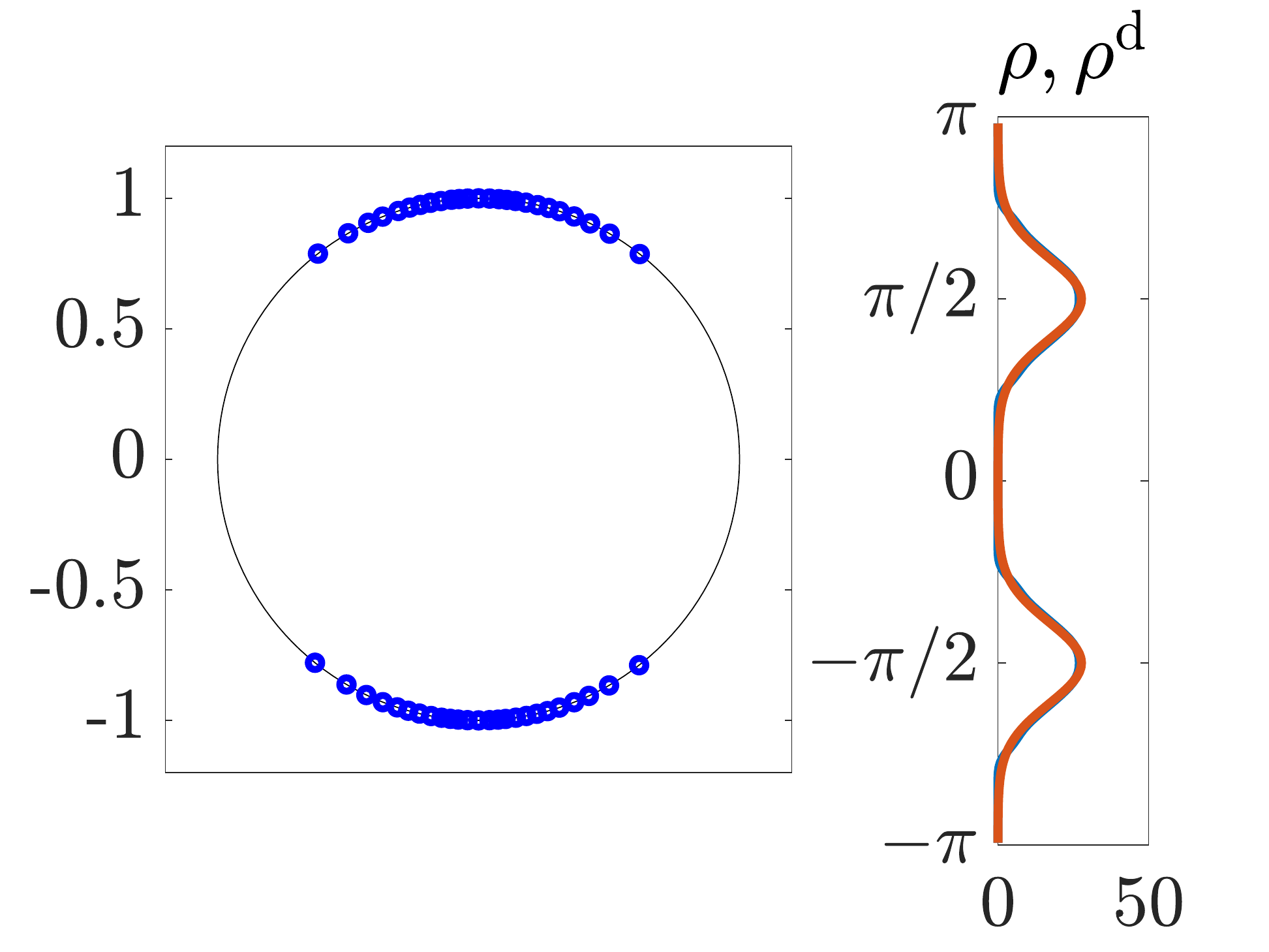}
         \caption{$t = 3$}
         \label{subfig::tf_bimpodal_reg}
     \end{subfigure}
     \begin{subfigure}[b]{0.24\textwidth}
         \centering
         \includegraphics[width=\textwidth]{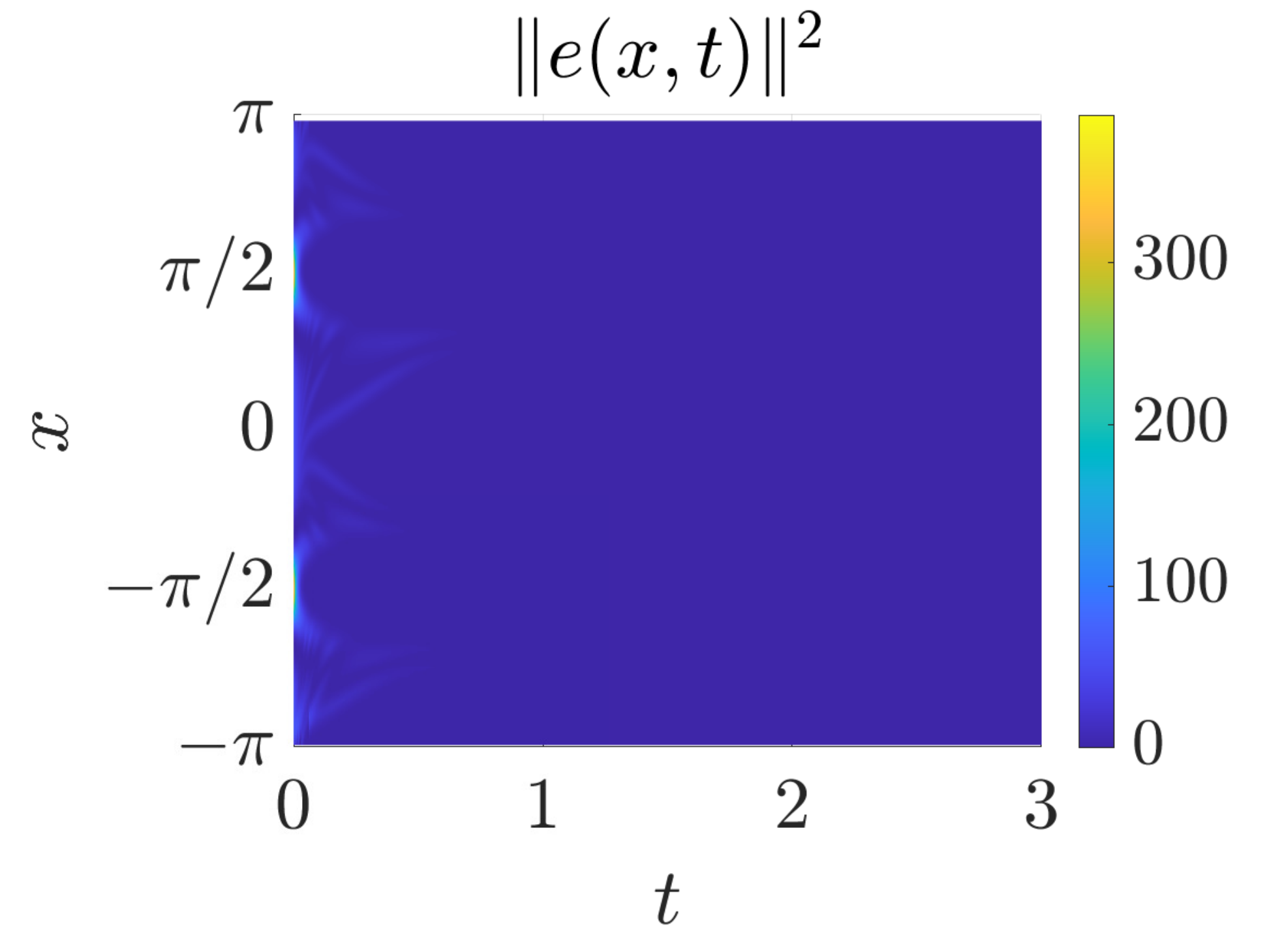}
         \caption{$\Vert e(x, t)\Vert$}
         \label{subfig::e_x_t_bimodal_reg}
     \end{subfigure}
     \begin{subfigure}[b]{0.24\textwidth}
         \centering
         \includegraphics[width=\textwidth]{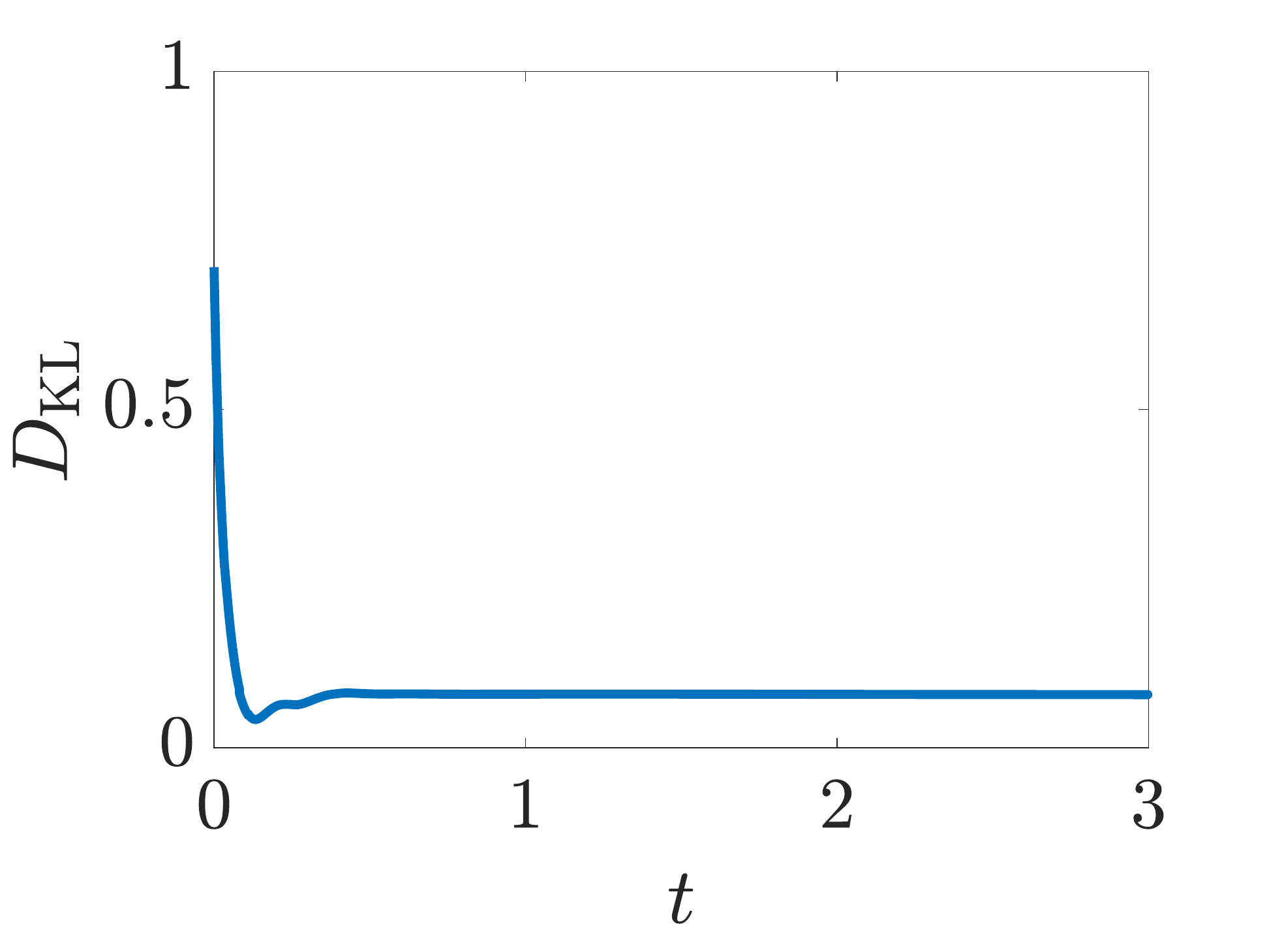}
         \caption{$D_{KL}(t)$}
         \label{subfig::D_KL_bimpodal_reg}
     \end{subfigure}
        \caption{Regulation to a bimodal configuration:(a) initial and (b) final configuration of the agents with their associated density (in blue) and desired density (in orange); (c) time and space evolution of the norm of the error function; (d) KL divergence between the desired normalised density and the normalised density.}
        \label{fig::bimpodal_reg}
\end{figure*}
\begin{figure}
     \centering
     \begin{subfigure}[b]{0.23\textwidth}
         \centering
         \includegraphics[width=\textwidth]{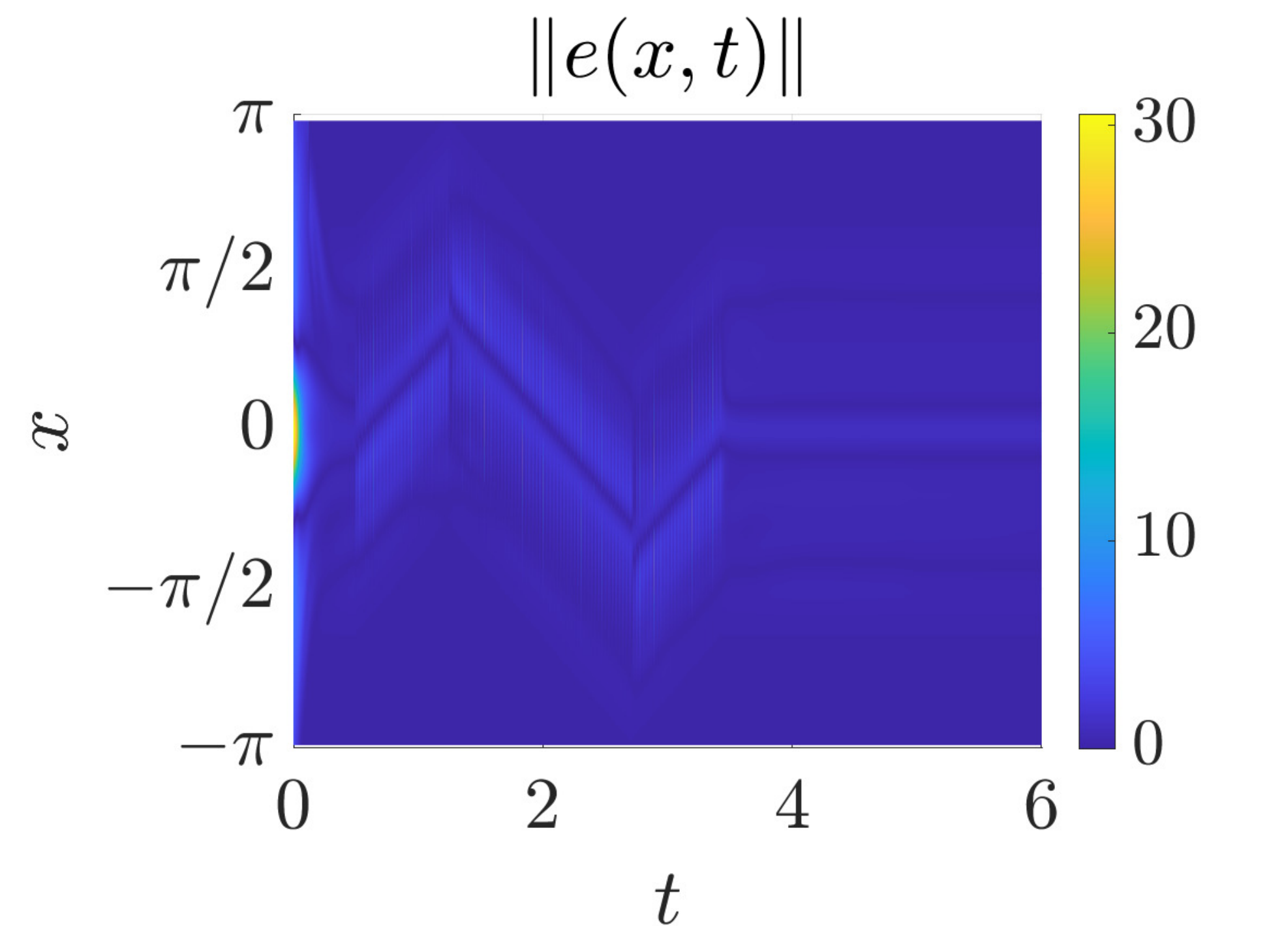}
         \caption{$\Vert e(x, t) \Vert$}
         \label{subfig::tracking_e_xt}
     \end{subfigure}
     \begin{subfigure}[b]{0.23\textwidth}
         \centering
         \includegraphics[width=\textwidth]{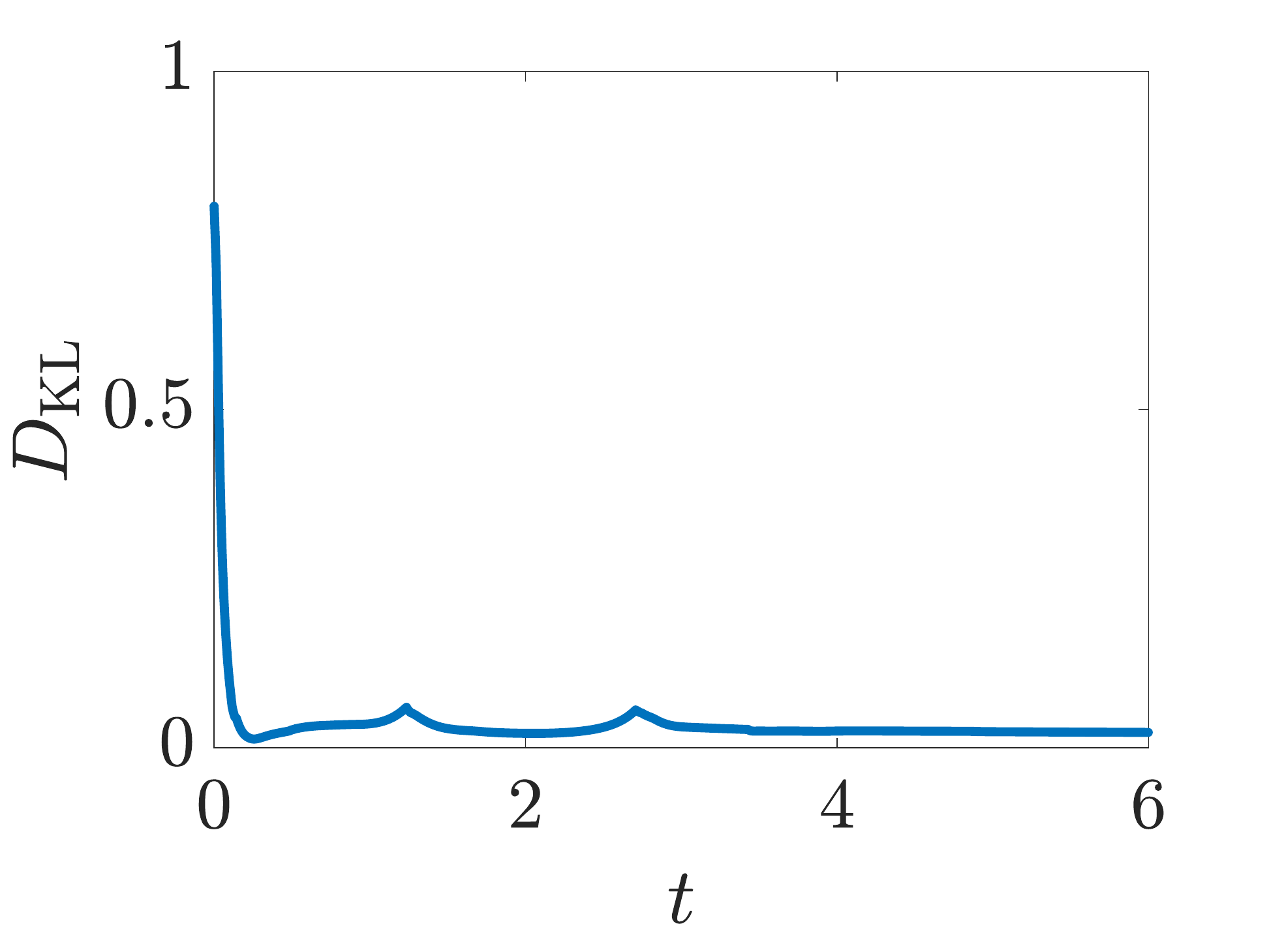}
         \caption{$D_{\mathrm{KL}}(t)$}
         \label{subfig::tracking_d_KL}
     \end{subfigure}
        \caption{Tracking a time-varying density: (a) time and space evolution of the norm of the error, (b) time evolution of the KL divergence.}
        \label{fig::tracking}
\end{figure}

Note that the feedback control action $q(x,t)$ consists of three terms. The first two terms are local control actions, while the latter is non-local, involving the convolution between the interaction kernel and the error. 
Such non-locality is practically mitigated by the assumption of considering a vanishing interaction kernel (see fig.~\ref{fig::forces}). 

\subsection{Discretization and microscopic control}
Next, we need to discretize the macroscopic control action in order to obtain the control inputs $u_i$  that can be deployed to steer the agents' microscopic dynamics \eqref{eq:themodel}.
Firstly, we recast the macroscopic controlled model in \eqref{eq:controlled_model} to include $q(x,t)$ as a control action on the velocity as
\begin{align}
    \rho_t(x,t)  + \left[\rho(x,t) (V(x,t) + U(x, t))\right]_x = 0,
\end{align}
where $U$ is an auxiliary function computed from the linear PDE (assuming $\rho(x,t)\ne 0$)
\begin{align}
    \left[\rho(x, t) U(x, t)\right]_x = -q(x, t).
    \label{eq:v}
\end{align}
The case $\rho(x,t)= 0$ corresponds to a case in which $q(x,t)$ is effectively behaving as a source/sink, changing the mass of the system (impossible without affecting the total number of agents in the system).
Integrating \eqref{eq:v}, we obtain
\begin{align}
    U(x, t) = -\frac{1}{\rho(x, t)}\left[\int_{-\pi}^x q(y, t)\,\mathrm{d}y + q(-\pi, t)\right].
\end{align}

To obtain the control action to be applied to the individual agent, we consider the agents as particles of a continuum, although not tagged with any label. We compute the velocity input acting on agent $i$, by sampling $U(x, t)$ at its position $x_i$, that is
\begin{align}
    u_i(t) = U(x_i, t), \quad i=1,2,\ldots N.
    \label{eq:u_i}
\end{align}

Note that (i) we are assuming a centralised scenario or that each agent either possesses or can estimate enough information about the other agents in the group in order to be able to locally compute $U(x_i,t)$. (ii) The assumption that $\rho(x,t)$ is nonzero is reasonable, as agents will estimate the density from their own positions, and hence we can choose a smoothing kernel such that $U(x, t)$ is always well-defined. Here, we use a Gaussian kernel estimation, adapted to take into account the domain's periodicity.
Moreover, since we are interested in discretizing the spatial control action, we know that $\rho(x, t)$ will be different from zero at least where there are effectively agents to control, that is, $U(x, t)$ is surely well-defined where we need to discretize it. (iii) The discretized controller will fulfill asymptotic convergence of agents' density to the desired one only when the number of agents is theoretically infinite. For any finite number of agents, convergence will be bounded and therefore \eqref{eq:controlgoal} remains satisfied. 

\section{Validation}
\label{sec:monomodal_regulation}
We validate the proposed strategy by selecting as a representative case of study the interaction kernel derived from a Morse potential, often used in the literature \cite{leverentz2009asymptotic, bernoff2011primer, mogilner2003mutual},
\begin{equation}
    f(z) = \mathrm{sgn}(z) \left[-G\text{e}^{-\vert z\vert / L} + \text{e}^{-\vert z\vert}\right],
\end{equation}
where, $G>0$ and  $L>0$ modulate the strength and characteristic distance of an attractive term, while the second term models repulsion normalized to have unitary repulsive strength and length scale as in \cite{leverentz2009asymptotic, bernoff2011primer}.
We choose $G$ and $L$ so that the repulsive interaction is dominant (as for example depicted in fig.~\ref{subfig::f2}). In particular we choose $G=L=0.5$.
We then address the problem of driving $N=50$ agents to converge towards different stable aggregation scenarios despite the fact that they would tend to repel each other away in the absence of control.  
We choose $K_\text{P} = 10$ and set the initial positions of the agents as evenly distributed in $\mathcal{S}$, $\rho^0(x) = N/2\pi$. We consider both regulation and tracking scenarios where agents need to converge towards a time-invariant or time-varying desired density profile. We also assess convergence and robustness of the proposed control strategy.


In order to quantify the {\em steady-state error} associated to a given trial, we use the Kullback-Leibler (KL) divergence \cite{kullback1951information} to estimate the distance between the desired density profile and the  density estimated from the positions of the agents, $D_{KL}(\hat{\rho}\Vert \hat{\rho}^\text{d})$, where $\hat{\rho}$ and $\hat{\rho}^\text{d}$ are the normalised versions of $\rho$ and $\rho^\text{d}$ such that their sum is 1. 


\subsection{Regulation}
Firstly, we consider a regulation task where agents should achieve a static desired density profile, $\rho^\text{d}(x, t) = \rho^\text{d}(x)$. We consider both a monomodal and a bimodal desired density.

As a first trial, we choose $\rho^\mathrm{d}$ as the following von Mises function \cite{mardia2000directional} with prescribed mean $\mu$ and  concentration coefficient $k$:
\begin{equation}
    \rho^\text{d} (x) = \frac{N\text{e}^{k\cos(x-\mu)}}{2\pi I_0(k)},
    \label{eq:vonmises}
\end{equation}
where $N$ is used to let the desired density sum to the total number of agents and $I_0$ is the modified Bessel function of the first kind of order 0 \cite{mardia2000directional}. We set $\mu = 0$ and $k = 4$.

Figures \ref{fig::monompodal_reg}(a)-(b) show the initial and final configuration of the agents and their associated density (compared with the desired one). The evolution of the error norm in space and time is depicted in fig. \ref{subfig::e_x_t_monomodal_reg}, while the time evolution of the KL divergence is shown in fig.~\ref{subfig::D_KL_monompodal_reg}. The results confirm the effectiveness of the proposed strategy with the control error converging quickly (in less than one time unit) to a small value and the agents achieving the desired configuration. 
We also show the control inputs at the microscopic level in fig. \ref{fig:control_input_monomodal}, resulting in signals converging to constant values. 
Note that the nonzero residual error shown in  fig.~\ref{subfig::D_KL_monompodal_reg} is an effect of the discretization of the macroscopic control action and does indeed converge to zero as the number of agents increases. 


As a further test, we also consider the problem of achieving a reference density which is the bimodal combination of two von Mises functions with the same concentration parameters and different means,
\begin{equation}
    \rho^\text{d} (x) = \frac{N}{4\pi I_0(k)} \left[\text{e}^{k\cos(x-\mu_1)} + \text{e}^{k\cos(x-\mu_2)}\right],
\end{equation}
where we  set $k = 8$ and $\mu_1 = \pi/2$ and  $\mu_2 = -\pi/2$. 
In fig.~\ref{fig::bimpodal_reg} we show the results of our simulations confirming the ability of the proposed strategy to achieve the desired control goal. In this trial, for brevity, we omit the evolution of the control inputs, which are qualitatively similar to those shown in fig.~\ref{fig:control_input_monomodal} for the previous case.

\subsection{Tracking}
To evaluate the ability of our strategy to track a time-varying desired density profile, we choose as target configuration a von Mises function such as \eqref{eq:vonmises} with a constant concentration parameter but a time-varying mean $\mu(t)$ which is 
null for the first 0.5s of the simulation and then increases with rate $\dot{\mu} = 1.47$rad/s, until reaching the value $\pi/3$. From that time instant, $\mu$ decreases with rate $\dot{\mu} = -1.47$rad/s until reaching the value $-\pi/3$, increasing again with rate $\dot{\mu} = 1.47$rad/s until returning to zero.
The evolution of the error norm in space and time and the KL divergence between $\rho$ and $\rho^\mathrm{d}$ is shown in fig.~\ref{fig::tracking}, confirming the viability of the proposed strategy and its effectiveness in steering agents' behavior towards the desired time-varying configuration. 

\subsection{Robustness}
\begin{figure}[t]
     \centering
     \begin{subfigure}[b]{0.24\textwidth}
         \centering
         \includegraphics[width=\textwidth]{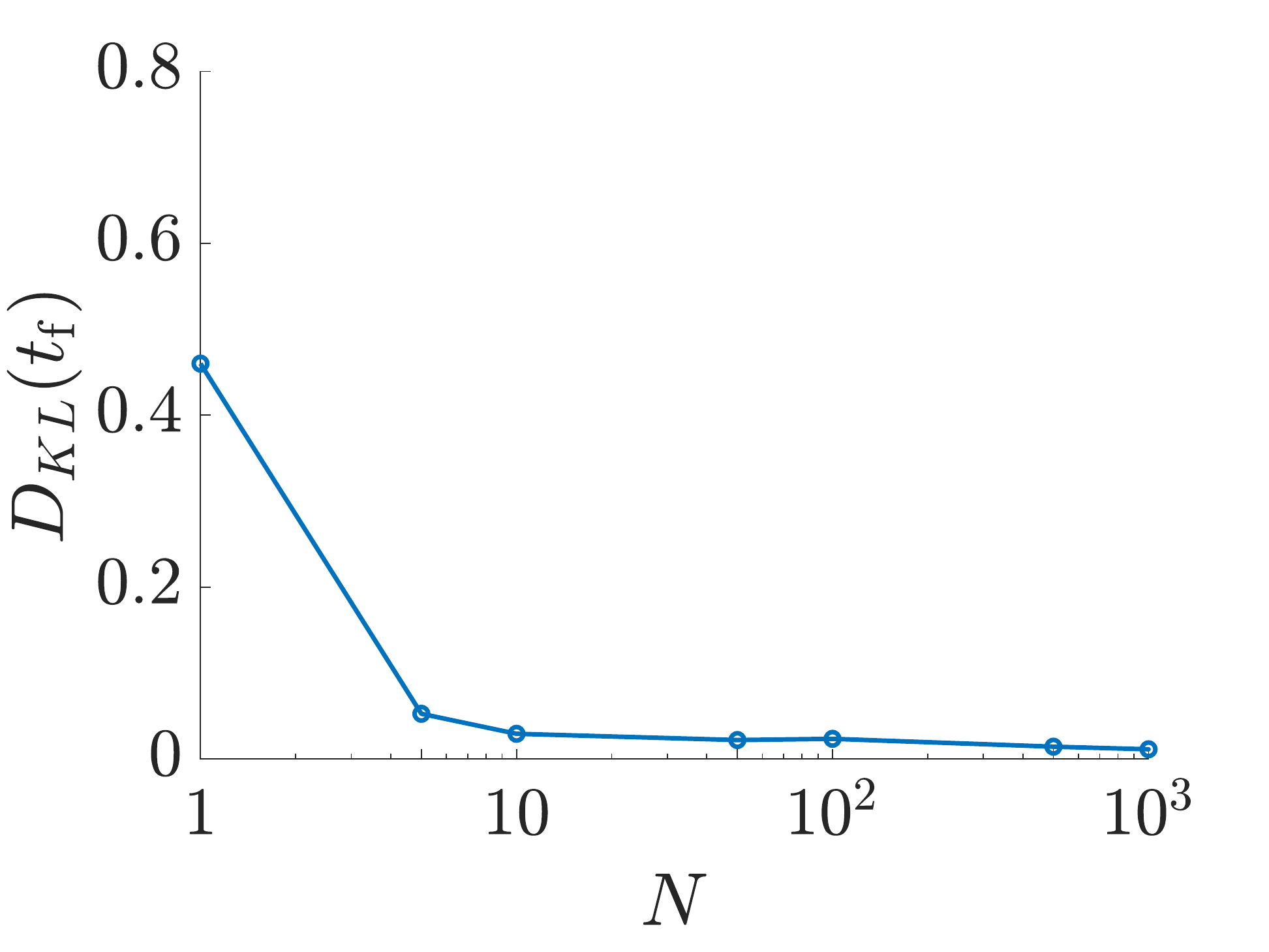}
         \caption{}
         \label{subfig::scalability}
     \end{subfigure}
     \begin{subfigure}[b]{0.23\textwidth}
         \centering
         \includegraphics[width=\textwidth]{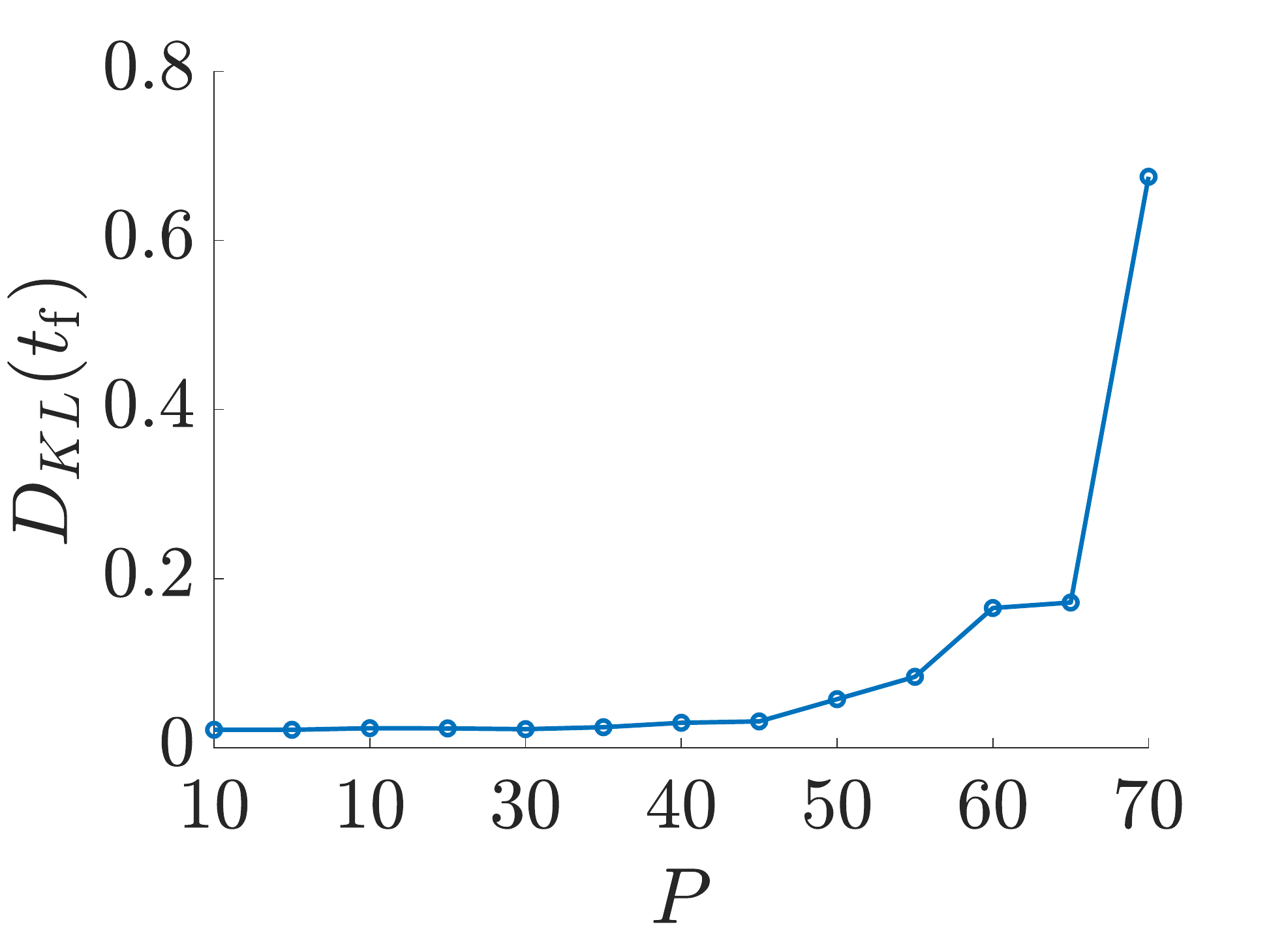}
         \caption{}
         \label{subfig::noise}
     \end{subfigure}
        \caption{Robustness tests: (a) scalability test - the KL divergence at the end of the trial is recorded for different values of $N$, (b) mesurement noise - the KL divergence at the end of the trial is recorded for different values of $P$ [dbW].}
        \label{fig::robustness}
\end{figure}
Finally, we test the robustness of the proposed strategy to changes in the number of agents and to measurement noise.

\subsubsection{Scalability}
To assess the scalability of the proposed control strategy to different numbers of agents, we run simulations for different values of $N$, and for each trial we record the value of $D_{KL}$ at the end of the simulation. We consider as test scenario the one discussed in sec.~\ref{sec:monomodal_regulation}. We report the results of such test in fig.~\ref{subfig::scalability}, for $N$ spanning from 1 to $10^3$. As the proposed strategy is based on a continuum approximation of the discrete set of agents of interest, we notice that, as expected, the steady state error becomes smaller as $N$ increases, becoming sufficiently smaller than 0.2 after $N$ gets larger than 5. Notice that the case $N=+\infty$ is reported as well in fig.~\ref{subfig::scalability}. This result was obtained performing a finite difference approximation of the continuified controlled model in \eqref{eq:controlled_model}. The numerical implementation was inspired by that reported in \cite{porfiri2007decline}.

\subsubsection{Measurement noise}
In order to assess the robustness of the proposed strategy, we perform additional simulations adding to \eqref{eq::q} some white noise with different power $P$ (measured in dbW). For each of these simulations, which use the same set-up considered in sec.~\ref{sec:monomodal_regulation}, we record the KL divergence at the end of the trials. Results are reported in fig.~\ref{subfig::noise}. We observe that, as the noise power increases past $P=60$ dbW, the steady-state mismatch between the agents' density and the desired one worsens. 

\section{Conclusions and Future Work}
We developed a continuification-based control strategy for a swarm of agents moving on a periodic bounded domain. We started by deriving a macroscopic model of agents' distribution on the ring and designed a  control action able to steer it to a desired configuration, proving its convergence. The microscopic control strategy was then obtained by spatially sampling the macroscopic control function at the agents' positions. Numerical simulations confirmed the effectiveness and robustness of the proposed approach. The extension to higher dimensional scenarios is the subject of ongoing work.







\bibliographystyle{IEEEtran}

\begin{thebibliography}{10}
\providecommand{\url}[1]{#1}
\csname url@samestyle\endcsname
\providecommand{\newblock}{\relax}
\providecommand{\bibinfo}[2]{#2}
\providecommand{\BIBentrySTDinterwordspacing}{\spaceskip=0pt\relax}
\providecommand{\BIBentryALTinterwordstretchfactor}{4}
\providecommand{\BIBentryALTinterwordspacing}{\spaceskip=\fontdimen2\font plus
\BIBentryALTinterwordstretchfactor\fontdimen3\font minus
  \fontdimen4\font\relax}
\providecommand{\BIBforeignlanguage}[2]{{%
\expandafter\ifx\csname l@#1\endcsname\relax
\typeout{** WARNING: IEEEtran.bst: No hyphenation pattern has been}%
\typeout{** loaded for the language `#1'. Using the pattern for}%
\typeout{** the default language instead.}%
\else
\language=\csname l@#1\endcsname
\fi
#2}}
\providecommand{\BIBdecl}{\relax}
\BIBdecl

\bibitem{Rubenstein2014}
M.~Rubenstein, A.~Cornejo, and R.~Nagpal, ``{Programmable self-assembly in a
  thousand-robot swarm},'' \emph{Science}, vol. 345, no. 6198, pp. 795--799,
  2014.

\bibitem{Gardi2022}
G.~Gardi, S.~Ceron, W.~Wang, K.~Petersen, and M.~Sitti, ``{Microrobot
  collectives with reconfigurable morphologies, behaviors, and functions},''
  \emph{Nature Communications}, vol.~13, no.~1, pp. 1--14, 2022.

\bibitem{giusti2022distributed}
A.~Giusti, G.~C. Maffettone, D.~Fiore, M.~Coraggio, and M.~di~Bernardo,
  ``Distributed control for geometric pattern formation of large-scale
  multirobot systems,'' \emph{arXiv preprint arXiv:2207.14567}, 2022.

\bibitem{guarino2020balancing}
A.~Guarino, D.~Fiore, D.~Salzano, and M.~di~Bernardo, ``Balancing cell
  populations endowed with a synthetic toggle switch via adaptive pulsatile
  feedback control,'' \emph{ACS Synthetic Biology}, vol.~9, no.~4, pp.
  793--803, 2020.

\bibitem{agrawal2019vitro}
D.~K. Agrawal, R.~Marshall, V.~Noireaux, and E.~D. Sontag, ``In vitro
  implementation of robust gene regulation in a synthetic biomolecular integral
  controller,'' \emph{Nature Communications}, vol.~10, no.~1, pp. 1--12, 2019.

\bibitem{calabrese2021spontaneous}
C.~Calabrese, M.~Lombardi, E.~Bollt, P.~De~Lellis, B.~G. Bardy, and
  M.~Di~Bernardo, ``Spontaneous emergence of leadership patterns drives
  synchronization in complex human networks,'' \emph{Scientific Reports},
  vol.~11, no.~1, pp. 1--12, 2021.

\bibitem{Shahal2020a}
S.~Shahal, A.~Wurzberg, I.~Sibony, H.~Duadi, E.~Shniderman, D.~Weymouth,
  N.~Davidson, and M.~Fridman, ``{Synchronization of complex human networks},''
  \emph{Nature Communications}, vol.~11, no.~1, pp. 1--10, 2020.

\bibitem{diBernardo2020}
M.~di~Bernardo, ``Controlling collective behavior in complex systems,'' in
  \emph{Encyclopedia of Systems and Control}, J.~Baillieul and T.~Samad,
  Eds.\hskip 1em plus 0.5em minus 0.4em\relax Springer London, 2020.

\bibitem{kardar2007statistical}
M.~Kardar, \emph{Statistical physics of particles}.\hskip 1em plus 0.5em minus
  0.4em\relax Cambridge University Press, 2007.

\bibitem{kardar2007statistical_fields}
------, \emph{Statistical physics of fields}.\hskip 1em plus 0.5em minus
  0.4em\relax Cambridge University Press, 2007.

\bibitem{Albi2020a}
G.~Albi, E.~Cristiani, L.~Pareschi, and D.~Peri, ``{Mathematical Models and
  Methods for Crowd Dynamics Control},'' \emph{Modeling and Simulation in
  Science, Engineering and Technology}, pp. 159--197, 2020.

\bibitem{Elamvazhuthi2021}
K.~Elamvazhuthi, Z.~Kakish, A.~Shirsat, and S.~Berman, ``{Controllability and
  Stabilization for Herding a Robotic Swarm Using a Leader: A Mean-Field
  Approach},'' \emph{IEEE Transactions on Robotics}, vol.~37, no.~2, pp.
  418--432, 2021.

\bibitem{Borzi2020}
A.~Borz{\`{i}} and L.~Gr{\"{u}}ne, ``{Towards a solution of mean-field control
  problems using model predictive control},'' \emph{Proc. 21st IFAC World
  Congress, Berlin, Germany}, vol.~53, no.~2, pp. 4973--4978, July 2020.

\bibitem{Kolpas2007}
A.~Kolpas, J.~Moehlis, and I.~G. Kevrekidis, ``{Coarse-grained analysis of
  stochasticity-induced switching between collective motion states},''
  \emph{Proceedings of the National Academy of Sciences}, vol. 104, no.~14, pp.
  5931--5935, 2007.

\bibitem{ascione2022mean}
G.~Ascione, D.~Castorina, and F.~Solombrino, ``Mean field sparse optimal
  control of systems with additive white noise,'' \emph{arXiv preprint
  arXiv:2204.02431}, 2022.

\bibitem{Gao2020}
S.~Gao and P.~E. Caines, ``{Graphon Control of Large-Scale Networks of Linear
  Systems},'' \emph{IEEE Transactions on Automatic Control}, vol.~65, no.~10,
  pp. 4090--4105, 2020.

\bibitem{lee2020coarse}
S.~Lee, M.~Kooshkbaghi, K.~Spiliotis, C.~I. Siettos, and I.~G. Kevrekidis,
  ``Coarse-scale {PDE}s from fine-scale observations via machine learning,''
  \emph{Chaos: An Interdisciplinary Journal of Nonlinear Science}, vol.~30,
  no.~1, p. 013141, 2020.

\bibitem{patsatzis2022data}
D.~G. Patsatzis, L.~Russo, I.~G. Kevrekidis, and C.~Siettos, ``Data-driven
  control of agent-based models: an equation/variable-free machine learning
  approach,'' \emph{arXiv preprint arXiv:2207.05779}, 2022.

\bibitem{nikitin2021continuation}
\BIBentryALTinterwordspacing
D.~Nikitin, C.~Canudas~de Wit, and P.~Frasca, ``A continuation method for
  large-scale modeling and control: from {ODE}s to {PDE}, a round trip,''
  \emph{IEEE Transactions on Automatic Control}, 2021. [Online]. Available:
  \url{10.1109/TAC.2021.3122387}
\BIBentrySTDinterwordspacing

\bibitem{Karafyllis2019}
I.~Karafyllis and M.~Papageorgiou, ``{Feedback control of scalar conservation
  laws with application to density control in freeways by means of variable
  speed limits},'' \emph{Automatica}, vol. 105, pp. 228--236, 2019.

\bibitem{Liard2020}
\BIBentryALTinterwordspacing
T.~Liard, R.~Stern, and M.~{Laura Delle Monache}, ``{A PDE-ODE model for
  traffic control with autonomous vehicles},'' 2020. [Online]. Available:
  \url{https://hal.archives-ouvertes.fr/hal-02492796}
\BIBentrySTDinterwordspacing

\bibitem{Zienkiewicz2018}
A.~K. Zienkiewicz, F.~Ladu, D.~A. Barton, M.~Porfiri, and M.~D. Bernardo,
  ``{Data-driven modelling of social forces and collective behaviour in
  zebrafish},'' \emph{Journal of Theoretical Biology}, vol. 443, pp. 39--51,
  2018.

\bibitem{Abaid2010}
N.~Abaid and M.~Porfiri, ``{Fish in a ring: Spatio-temporal pattern formation
  in one-dimensional animal groups},'' \emph{Journal of the Royal Society
  Interface}, vol.~7, no.~51, pp. 1441--1453, 2010.

\bibitem{DeLellis2020}
P.~{De Lellis}, E.~Cadolini, A.~Croce, Y.~Yang, M.~{Di Bernardo}, and
  M.~Porfiri, ``{Model-Based Feedback Control of Live Zebrafish Behavior via
  Interaction with a Robotic Replica},'' \emph{IEEE Transactions on Robotics},
  vol.~36, no.~1, pp. 28--41, 2020.

\bibitem{Aureli2010}
M.~Aureli and M.~Porfiri, ``{Coordination of self-propelled particles through
  external leadership},'' \emph{Europhysics Letters}, vol.~92, no.~4, 2010.

\bibitem{Yates2009}
C.~A. Yates, R.~Erban, C.~Escudero, I.~D. Couzin, J.~Buhl, I.~G. Kevrekidis,
  P.~K. Maini, and D.~J. Sumpter, ``{Inherent noise can facilitate coherence in
  collective swarm motion},'' \emph{PNAS}, vol. 106, no.~14, pp. 5464--5469,
  2009.

\bibitem{leverentz2009asymptotic}
A.~J. Leverentz, C.~M. Topaz, and A.~J. Bernoff, ``Asymptotic dynamics of
  attractive-repulsive swarms,'' \emph{SIAM Journal on Applied Dynamical
  Systems}, vol.~8, no.~3, pp. 880--908, 2009.

\bibitem{bernoff2011primer}
A.~J. Bernoff and C.~M. Topaz, ``A primer of swarm equilibria,'' \emph{SIAM
  Journal on Applied Dynamical Systems}, vol.~10, no.~1, pp. 212--250, 2011.

\bibitem{viscek1995}
T.~Viscek, A.~Czir{\`{o}}k, E.~Ben-Jacob, I.~Cohen, and O.~Shochet, ``{Novel
  Type of Phase Transition in a System of Self-Driven Particles},''
  \emph{Physical Review Letters}, vol.~75, no.~6, pp. 1226--1229, 1995.

\bibitem{Bodnar2005a}
M.~Bodnar and J.~J. Velazquez, ``{Derivation of macroscopic equations for
  individual cell-based models: A formal approach},'' \emph{Mathematical
  Methods in the Applied Sciences}, vol.~28, no.~15, pp. 1757--1779, 2005.

\bibitem{young1912multiplication}
W.~H. Young, ``On the multiplication of successions of fourier constants,''
  \emph{Proc. Roy. Soc. A}, vol.~87, no. 596, pp. 331--339, 1912.

\bibitem{mogilner2003mutual}
A.~Mogilner, L.~Edelstein-Keshet, L.~Bent, and A.~Spiros, ``Mutual
  interactions, potentials, and individual distance in a social aggregation,''
  \emph{Journal of Mathematical Biology}, vol.~47, no.~4, pp. 353--389, 2003.

\bibitem{kullback1951information}
S.~Kullback and R.~A. Leibler, ``On information and sufficiency,'' \emph{The
  Annals of Mathematical Statistics}, vol.~22, no.~1, pp. 79--86, 1951.

\bibitem{mardia2000directional}
K.~V. Mardia, P.~E. Jupp, and K.~Mardia, \emph{Directional statistics}.\hskip
  1em plus 0.5em minus 0.4em\relax Wiley Online Library, 2000, vol.~2.

\bibitem{porfiri2007decline}
M.~Porfiri, E.~M. Bollt, and D.~J. Stilwell, ``Decline of minorities in
  stubborn societies,'' \emph{The European Physical Journal B}, vol.~57, no.~4,
  pp. 481--486, 2007.

\end{thebibliography}

\end{document}